\documentclass[10pt,journal,comsoc]{IEEEtran}  

\usepackage{cite}
\usepackage{amsmath,amssymb,amsfonts}
\usepackage{graphicx}
\usepackage{graphics}
\usepackage{epstopdf} 
\usepackage{subfigure}
\usepackage[linesnumbered,ruled,vlined,algo2e]{algorithm2e}
\usepackage{multirow}
\usepackage{array}
\usepackage{textcomp}
\usepackage{xcolor}
\usepackage{multicol}
\usepackage{makecell}
\usepackage{bm}
\usepackage{tikz}
\usetikzlibrary{shapes.geometric, arrows, positioning, calc, fit}

\usepackage{amsthm}

\def\BibTeX{{\rm B\kern-.05em{\sc i\kern-.025em b}\kern-.08em
		T\kern-.1667em\lower.7ex\hbox{E}\kern-.125emX}}

\begin{document}

\title{Joint Energy Efficiency and Fairness Optimization for D2D Communications in Aerial-Ground Integrated Heterogeneous Networks}


\author{Chuan-Chi~Lai,~\IEEEmembership{Member,~IEEE},~Ang-Hsun~Tsai,~\IEEEmembership{Member,~IEEE},~and~Shang-Long Wu%
\color{black}
	\IEEEcompsocitemizethanks{		
		\IEEEcompsocthanksitem Copyright (c) 2026 IEEE. Personal use of this material is permitted. However, permission to use this material for any other purposes must be obtained from the IEEE by sending a request to pubs-permissions@ieee.org.
		\IEEEcompsocthanksitem This research was supported in part by the National Science and Technology Council, Taiwan, R.O.C., under Grant Nos. \mbox{NSTC 114-2221-E-194-062-,} NSTC 115-2221-E-194-042-MY2, and \mbox{NSTC 115-2221-E-035-050-.} This work was also supported in part by the Advanced Institute of Manufacturing with High-tech Innovations (AIM-HI) from the Featured Areas Research Center Program within the framework of the Higher Education Sprout Project by the Ministry of Education (MOE) in Taiwan. In addition, this work was sponsored in part by Feng Chia University under Grant 25H00812.
		\emph{(Corresponding author: Ang-Hsun~Tsai.)}
		\IEEEcompsocthanksitem C.-C. Lai is with the Department of Communications Engineering, National Chung Cheng University, Minxiong Township, Chiayi County 621301, Taiwan, and also with the Advanced Institute of Manufacturing with High-tech Innovations (AIM-HI), National Chung Cheng University, Minxiong Township, Chiayi County 621301, Taiwan (e-mail: chuanclai@ccu.edu.tw). 
		\IEEEcompsocthanksitem A.-H. Tsai is with the Department of Communications Engineering, Feng Chia University, Taichung 407102, Taiwan (email:ahtsai@fcu.edu.tw).
		\IEEEcompsocthanksitem S.-L. Wu is with the Department of Communications Engineering, National Chung Cheng University, Minxiong Township, Chiayi County 621301, Taiwan. 
	}
}


\maketitle

\begin{abstract}
This study investigates an Aerial-Ground Integrated Heterogeneous network (AGIHN) architecture that combines terrestrial macro base stations and unmanned aerial vehicles (UAVs) serving as aerial base stations to enhance uplink access for macrocell users. To address the complex uplink resource allocation challenge for multiple device-to-device (D2D) communication pairs, we propose a low-complexity Multi-Channel Rate-Fair (MCRF) algorithm. Distinct from traditional exclusive allocation methods, MCRF supports shared reuse, enabling multiple D2D pairs to simultaneously multiplex on the same resource block, thereby significantly improving spectral efficiency. To manage the severe intra-tier interference arising from this non-orthogonal sharing, a heuristic Interference Avoidance (IA) strategy is integrated to ensure the transmission quality of D2D users. The proposed framework jointly optimizes system throughput, user fairness, and energy efficiency without requiring computationally intensive offline training. Simulation results demonstrate distinct performance advantages depending on the reuse mode: Compared to traditional single-channel exclusive reuse schemes, MCRF achieves massive gains, increasing D2D energy efficiency and throughput by approximately 397\% and 542\%, respectively. Furthermore, relative to multi-channel benchmarks (e.g., MCRR), the proposed algorithm optimizes the efficiency-fairness trade-off, enhancing the fairness index by 7.43\% while maintaining a robust fairness score exceeding 0.6 in interference-prone environments.
\end{abstract}


\section{Introduction}
\label{sec:introduction}
\IEEEPARstart{I}{n} recent years, the rapid evolution of wireless broadband and mobile communications has revolutionized everyday life through ubiquitous Internet access. However, spectrum scarcity remains a critical constraint due to the soaring demand from smartphones and rich multimedia services, which often leads to congestion and unstable signals in crowded environments~\cite{Nagarajan2025,7018136,10158439}. While deploying additional terrestrial base stations can provide partial relief, the 3GPP has proposed proactive enhancements in LTE-Advanced such as device-to-device (D2D) communication to significantly improve spectral efficiency~\cite{3gpp-tr-22.803,TR36.843}.

D2D communication enables direct links between nearby user devices under the supervision of a cellular network, eliminating the need for base station relay~\cite{7128330}. This approach improves spectrum utilization, increases system throughput, and enhances energy efficiency through reduced transmission power~\cite{10184474,10423373,8114722,9072416}. Embedded in 3GPP standards, D2D technology has been extensively studied with a focus on device discovery, channel modeling, and both in-coverage and out-of-coverage scenarios~\cite{TR36.843}.

Resource management for D2D is generally classified into centralized and distributed models. In centralized control, the base station allocates resources in underlay mode, where D2D pairs share spectrum with cellular users and require interference mitigation, or in overlay mode, where a dedicated portion of the spectrum is reserved for D2D at the cost of reduced reuse. In distributed control, devices manage access autonomously, often over unlicensed bands, and employ decentralized protocols such as listen-before-talk. Innovations such as 5G-Advanced sidelink over unlicensed spectrum (SL-U) demonstrate the potential of distributed operation in dense deployments~\cite{yan2025performance,FRANGULEA2025108143}.

In conventional D2D networks, throughput is strongly influenced by resource allocation strategies. Most centralized approaches limit each uplink resource block (RB) to a single D2D pair, simplifying interference management but reducing spectrum utilization. Recent surveys~\cite{10.4108/eai.4-5-2022.173977,ISLAM2022102978} indicate that allowing multiple D2D pairs to share the same RB can significantly enhance throughput and fairness by exploiting multiuser diversity.

In addition, a recent and promising trend is the integration of Aerial-Ground Integrated Heterogeneous Networks (AGIHNs), which combine unmanned aerial vehicle base stations (UAV-BSs) with terrestrial infrastructure. This architecture offers agile and on-demand coverage in scenarios such as disaster recovery, temporary large-scale events, and rural connectivity. In AGIHNs, D2D can offload traffic from both aerial and ground infrastructure, reduce latency through proximity-based relaying, and extend connectivity to edge users. However, the coexistence of aerial and ground tiers introduces challenges including dynamic topology changes, complex interference patterns, and time-varying link qualities, which make resource allocation significantly more complex~\cite{Zhao2024UAVD2D,10278101,Deng2025}.

Motivated by the above studies, this work proposes a novel resource allocation framework for uplink D2D communication that permits multiple D2D pairs to reuse the same RB. The design addresses both traditional terrestrial networks and AGIHNs, taking into account aerial-ground integration, dynamic network topologies, heterogeneous link qualities, and cross-tier interference mitigation. The objective is to improve fairness among D2D users, enhance energy efficiency, and protect the performance of cellular transmissions, contributing to the development of highly adaptive and spectrum-efficient D2D-enabled heterogeneous networks.

\begin{figure}
	\centering
	\includegraphics[width=\columnwidth]{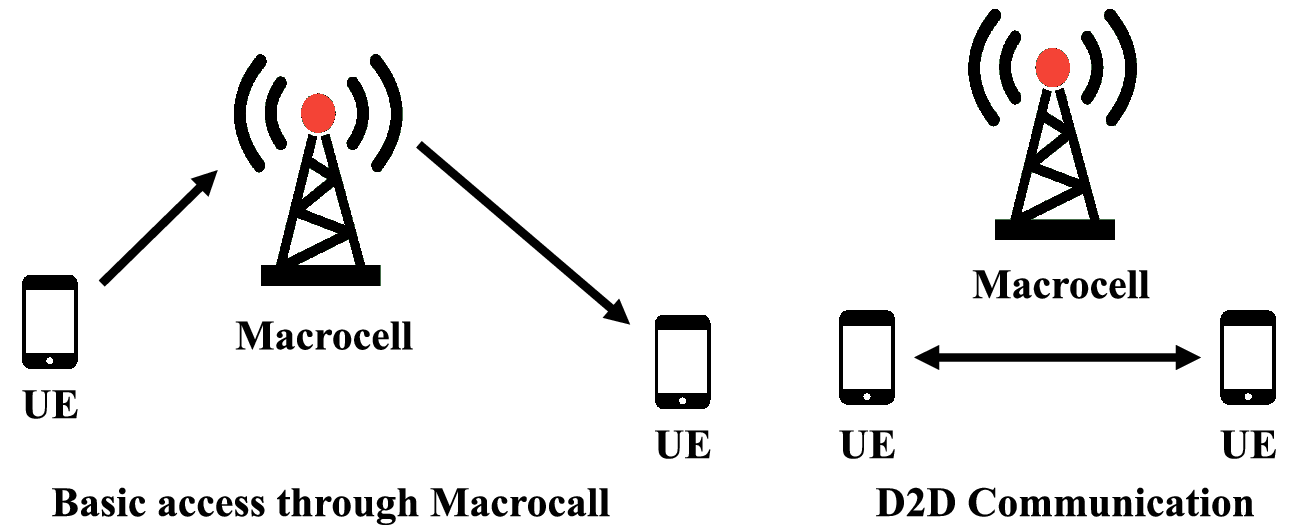}
	\caption{Cellular Communication v.s. D2D Communication.}
	\label{fig:fig1.pdf}
\end{figure}

Hence, the contributions of this work are summarized as follows:
\begin{itemize}
    \item We formulate a joint energy efficiency and fairness optimization problem for uplink transmission in an Air-Ground Integrated Heterogeneous Network (AGIHN). Unlike traditional approaches, we specifically address the challenging non-orthogonal shared reuse scenario, where multiple D2D pairs simultaneously multiplex on the same resource blocks, introducing severe intra-tier interference.
    
    \item To solve this NP-hard problem without the high computational overhead of learning-based methods, we propose a low-complexity, training-free Multi-Channel Rate-Fair (MCRF) algorithm. This approach integrates a fairness-aware heuristic with a sophisticated Interference Avoidance (IA) strategy to efficiently manage radio resources in dynamic environments.
    
    \item The proposed framework implements a robust interference management strategy that acts as a ``spectral gatekeeper.'' It effectively improves the system throughput and energy efficiency of D2D users by exploiting spatial multiplexing gains, while strictly protecting the Quality of Service (QoS) of primary cellular users (MUEs) from aggregate interference.
    
    \item Simulation results validate the distinct advantages of the proposed strategy depending on the reuse mode. Compared to traditional exclusive reuse (single-channel) schemes, the MCRF algorithm utilizes spectral aggregation to improve D2D Energy Efficiency and Throughput by approximately 397\% and 542\%, respectively. Furthermore, relative to multi-channel benchmarks, it optimizes the efficiency-fairness trade-off, enhancing the Fairness Index by 7.43\%.
\end{itemize}

\color{black}

Section~\ref{sec:related_work} presents the related work. Section~\ref{sec:system_Model} introduces the system model. Section~\ref{sec:problem_formulation} describes the problem formulation. Section~\ref{sec:proposed_method} explains the proposed approach. The simulation results and discussion are presented in Section~\ref{sec:simulation}. Finally, Section~\ref{sec:conclusion} presents the conclusion of this study.

\section{Related Work}
\label{sec:related_work}

The evolution of resource allocation strategies has progressed from traditional terrestrial D2D networks to complex Aerial-Ground Integrated Heterogeneous Networks (AGIHNs). This section categorizes existing literature into terrestrial approaches, UAV-assisted management, and the specific challenges regarding fairness in 3D network architectures.

\subsection{Resource Allocation in Terrestrial D2D Networks}
Early research on D2D communications primarily focused on terrestrial cellular networks, establishing the fundamental principles of mode selection and spectrum reuse. In underlay scenarios, resource allocation typically aims to manage interference between Cellular Users (CUs) and D2D users while optimizing Energy Efficiency (EE) or spectral efficiency.

The investigation in~\cite{7880995} focused on uplink underlay modes where resource allocation followed the Max-Min Fairness (MMF) principle. The analysis highlighted that system EE significantly degrades as the distance between D2D pairs increases or as the number of shared Resource Blocks (RBs) decreases. Similarly, the study in~\cite{7938308} analyzed the impact of user distance on EE through minimum power control for MUEs, confirming the inverse relationship between D2D distance and system efficiency.

Regarding allocation principles, \cite{7564170} evaluated four distinct strategies: Proportional Fairness (PF), minimum transmission power, randomness, and minimum channel choice. The results demonstrated that while minimum power ensures the lowest consumption, PF maximizes system capacity while maintaining basic communication quality. This superiority of PF scheduling was further corroborated by~\cite{7881210} and~\cite{1545851} in multi-user OFDM systems, showing that controlling capacity ratios can effectively balance overall throughput and individual user data rates.

However, these conventional terrestrial approaches often simplify the optimization problem. For instance, \cite{7794815} utilized Karush-Kuhn-Tucker (KKT) conditions to optimize EE in small cell networks. While effective in 2D static environments, these methods typically restrict each RB to a single D2D pair or neglect the complex Line-of-Sight (LoS) interference patterns inherent in 3D network topologies.

\begin{table*}[t]
\centering
\caption{Qualitative Comparison of the Proposed Scheme with Existing Approaches}
\label{tab:comparison}
\resizebox{\textwidth}{!}{
\begin{tabular}{|c|c|c|c|c|c|c|c|c|}
\hline
\textbf{Reference} & \textbf{Year} & \textbf{Methodology} & \textbf{Network Scenario} & \textbf{Objective} & \textbf{\makecell{RB Reuse \\ Policy}} & \textbf{\makecell{Interference \\ Mitigation}} & \textbf{Fairness} & \textbf{\makecell{Computational \\ Complexity}} \\ \hline
\makecell{Shen \\ et al. \cite{1545851}} & 2005 & \makecell{Optimization \\ (Suboptimal)} & \makecell{MU-OFDM \\ (Downlink)} & \makecell{Sum \\ Capacity} & \makecell{Orthogonal \\ (1 User/RB)} & \makecell{Subchannel \\ Assignment} & \makecell{Yes \\ (Rate Constr.)} & Low \\ \hline
\makecell{Hu \\ et al. \cite{7938308}} & 2017 & \makecell{Iterative Algorithm \\ (Matching Theory)} & \makecell{D2D Underlay \\ (Downlink)} & \makecell{Energy \\ Efficiency} & \makecell{Exclusive \\ (1 D2D/RB)} & \makecell{Power Control \\ \& Matching} & \makecell{No \\ (QoS Only)} & Moderate \\ \hline
\makecell{Xu \\ et al. \cite{9415745}} & 2021 & \makecell{Learning \\ (LSTM/GAN)} & \makecell{UAV-assisted \\ Network} & \makecell{Traffic \\ Prediction} & N/A & N/A & No & \makecell{High \\ (Training Required)} \\ \hline
\makecell{Rahim \\ et al. \cite{Rahim2025}} & 2025 & \makecell{Optimization \\ (Guard Zone)} & \makecell{Terrestrial D2D \\ (Downlink)} & \makecell{Coverage \\ Probability} & \makecell{Spatial Reuse \\ (Guard Zone)} & \makecell{Guard Zone \\ \& Power Control} & No & Low \\ \hline
\makecell{Noman \\ et al. \cite{Noman2025}} & 2025 & \makecell{Learning \\ (Federated DDQN)} & D2D HetNets & \makecell{Energy \\ Efficiency} & \makecell{Shared \\ (Multiple D2Ds)} & \makecell{Joint Resource \\ Allocation} & \makecell{Yes \\ (QoS Only)} & \makecell{High \\ (Training Required)} \\ \hline
\textbf{\makecell{Proposed \\ MCRF}} & \textbf{2026} & \textbf{\makecell{Heuristic \\ Algorithm}} & \textbf{\makecell{3D UAV-AGIHN \\ (Uplink)}} & \textbf{\makecell{EE \& \\ Fairness}} & \textbf{\makecell{Shared \\ (Multiple D2Ds)}} & \textbf{\makecell{Interference \\ Avoidance (IA)}} & \textbf{\makecell{Yes \\ (Jain's Index)}} & \textbf{\makecell{Low \\ (Polynomial)}} \\ \hline
\end{tabular}
}
\end{table*}

\subsection{Resource Management in UAV-assisted and AGIHN Networks}
The integration of UAVs introduces new degrees of freedom and challenges, necessitating more advanced resource management strategies. Recent studies have expanded allocation frameworks to AGIHNs, focusing on the unique attributes of Air-to-Ground channels.

Several works have employed game-theoretic approaches to handle the interactions between aerial and ground nodes. Energy efficiency in UAV-underlay networks was addressed in~\cite{Zhao2024UAVD2D} using a Nash game for power control combined with the Hungarian algorithm for channel matching. Similarly, spectrum leasing in UAV-enabled Mobile Edge Computing (MEC) systems was explored in~\cite{Deng2025} using Stackelberg game theory. To address multi-objective optimization, \cite{10278101} jointly optimized UAV deployment, transmission power, and flight velocity.

Parallel to optimization-based methods, learning-based approaches have gained traction. A framework based on Generative Adversarial Networks (GAN) and LSTM was proposed in~\cite{9415745} for UAV-M2M communications to predict network demand and optimize resources. While Deep Reinforcement Learning (DRL) and similar techniques offer powerful solutions for dynamic environments, they often require substantial training data and computational overhead. These requirements may be prohibitive for rapid deployment in emergency or temporary AGIHN scenarios.

Furthermore, a Cooperative Distributed Resource Allocation Algorithm (CDRAA) was proposed in~\cite{10168283} for heterogeneous networks using a Time-Sharing Factor (TSF) technique. Although this improves throughput compared to centralized optimization, it does not fully account for the severe three-dimensional aerial-ground interference present when multiple UAVs and D2D pairs coexist.

\subsection{Fairness and Interference Mitigation in Shared Spectrum}
Ensuring fairness among D2D users sharing scarce spectrum remains a critical challenge in AGIHNs. While the Proportional Fairness (PF) principle has a solid theoretical foundation in cellular networks~\cite{1545851}, its direct application to AGIHNs faces limitations. In a 3D architecture, the interference from ground D2D transmitters to UAV uplink channels, and vice versa, is significantly more severe than in terrestrial networks due to the high probability of LoS links. Although recent studies like~\cite{Rahim2025} have proposed effective interference mitigation techniques such as dynamic guard zones, these are typically designed for terrestrial downlink scenarios and may not fully address the 3D mobility challenges in UAV-assisted networks.

Traditional PF schemes often fail to maintain SINR stability when multiple D2D pairs reuse the same RB under such aggressive interference. Additionally, the rapid channel state variations caused by UAV mobility make it difficult for existing frameworks to adapt in real-time, often leading to SINR dropping below the threshold. To tackle these dynamic challenges, recent works~\cite{Noman2025,9415745} have introduced Federated Multi-agent Deep Reinforcement Learning (F-MADDQN) frameworks. However, such data-driven approaches typically incur high computational overhead and require extensive training periods, which may not be suitable for time-sensitive disaster relief missions. These gaps highlight the need for a low-complexity heuristic that can balance system throughput and user fairness without the heavy training burden of learning-based methods.

Therefore, this study proposes a Multi-Channel Rate-Fair (MCRF) algorithm coupled with an Interference Avoidance (IA) strategy, specifically designed to address the trade-off between energy efficiency and fairness in high-density and interference-prone AGIHN environments.
To clearly position our contributions within the literature, Table \ref{tab:comparison} qualitatively contrasts the proposed scheme with state-of-the-art approaches.
A key distinction lies in the RB Reuse Policy. While traditional optimization methods \cite{7938308} typically enforce exclusive reuse (one D2D link per RB) and recent spatial-based schemes \cite{Rahim2025} rely on geographic guard zones, our framework supports shared reuse (multiple D2D links per RB). This allows for significantly higher spectral efficiency but necessitates the sophisticated IA strategy we propose to manage the resulting aggregate interference in 3D UAV environments.

\color{black}

\section{System Model} 
\label{sec:system_Model}

\subsection{System Architecture} 

\begin{figure}[!t]
	\centering
	\includegraphics[width=\columnwidth]{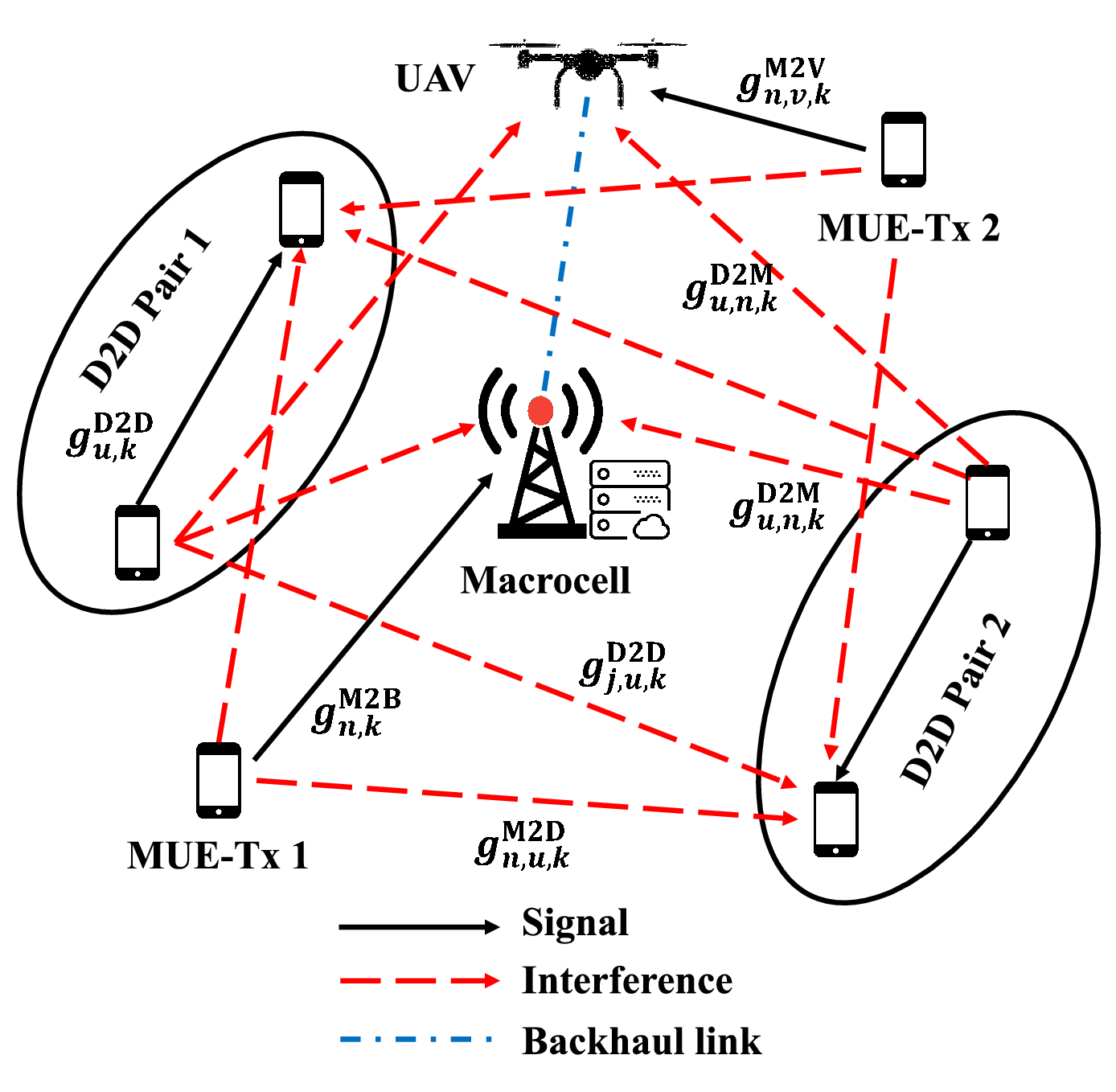}
	\caption{System model considered in this work.}
	\label{fig:system_model}
\end{figure}

This study is based on the Air-Ground Integrated Heterogeneous Network (AGIHN) architecture, utilizing an Orthogonal Frequency Division Multiplexing (OFDM) system with uplink transmission and adopting the underlay mode for resource allocation management.
The overall scenario includes one terrestrial macro base station (MBS) and $N_{\rm V}$ unmanned aerial vehicles (UAVs) acting as aerial base stations, jointly serving as uplink access points for macrocell users (MUEs). 
The considered network architecture is shown in Fig.~\ref{fig:system_model}, and the important symbols used in this study are summarized in Table~\ref{tab:mcrf_symbols}.

Each UAV establishes a backhaul link to the MBS, which is equipped with an edge controller responsible for centralized management and coordination of the entire network. The spectrum of the backhaul links from UAVs to the MBS and the fronthaul links from MUEs to each UAV are assumed to be orthogonal, ensuring no mutual interference between these links. This orthogonality allows reliable communication and simplifies resource allocation for both aerial and terrestrial segments of the network.

\vspace{0.5em} 
\noindent \textbf{Mobility and Quasi-Static Assumption:}
In the proposed AGIHN architecture, the UAV functions as an aerial base station. To provide stable and reliable coverage, the UAV is assumed to operate in a \textit{hovering or quasi-stationary state} during the service period. The core focus of this study is the resource allocation for ground D2D users rather than UAV trajectory optimization. 
For the mobile D2D pairs and MUEs on the ground, the time scale of resource allocation (typically $1$ ms per scheduling slot) is significantly shorter than their mechanical mobility scale. Even for vehicular users moving at high speeds (e.g., $30$ m/s), the maximum physical displacement within a single $1$ ms slot is merely $0.03$ m. Because this microscopic displacement does not induce noticeable changes in large-scale path loss or macro-level interference topology, we adopt a \textit{quasi-static assumption}. Under this assumption, the user positions are considered constant within a single scheduling time slot, and the proposed MCRF algorithm executes instantly on a snapshot basis to optimize resources for the current topology. This design ensures the system can adapt to mobility by continuously refreshing the allocation matrix frame-by-frame.

\begin{table}[!t]
	\renewcommand{\arraystretch}{1.2}
	\centering
	\caption{Summary of Symbols in This Study}
	\label{tab:mcrf_symbols}
	\begin{tabular}{|c|m{.775\columnwidth}|}
	\hline
	\textbf{Symbol} & \textbf{Description} \\ \hline
	$N_{\rm M}$ & Number of MUEs \\ \hline
	$N_{\rm D}$ & Number of D2D users \\ \hline
	$N_{\rm V}$ & Number of UAV base stations \\ \hline
	$N_0$ & Background noise \\ \hline
	$W_{\rm RB}$ & Bandwidth of each resource block\\ \hline
	$K$ & Number of resource blocks \\ \hline	
	$\alpha_{u,k}$ & Allocation status of the $k$-th resource block to the $u$-th D2D user pair\\ \hline
	$\alpha_{n,k}$ & Allocation status of the $k$-th resource block to the $n$-th MUE\\ \hline
	$\delta_{n,\rm B}$ & Association between the $n$-th MUE and the MBS\\ \hline
	$\delta_{n,v}$ & Association between the $n$-th MUE and the $v$-th UAV\\ \hline
	$P_{u,k}^{\rm D}$ & Transmit power of the $u$-th D2D pair on the $k$-th resource block\\ \hline
	$P_{n,k}^{\rm M}$ & Transmit power of the $n$-th MUE on the $k$-th resource block\\ \hline
	$g_{n,k}^{\rm M2B}$ & Channel gain between the $n$-th MUE and the MBS on the $k$-th resource block\\ \hline
	$g_{n,u,k}^{\rm M2D}$ & Channel gain between the $n$-th MUE and the receiver of the $u$-th D2D pair on the $k$-th resource block\\ \hline
	$g_{n,v,k}^{\rm M2V}$ & Channel gain between the $n$-th MUE and the $v$-th UAV on the $k$-th resource block\\ \hline
	$g_{u,k}^{\rm D2B}$ & Channel gain between the transmitter of the $u$-th D2D pair and the MBS on the $k$-th resource block\\ \hline
	$g_{u,v,k}^{\rm D2V}$ & Channel gain between the transmitter of the $u$-th D2D pair and the $v$-th UAV on the $k$-th resource block\\ \hline
	$g_{u,k}^{\rm D2D}$ & Channel gain between the transmitter and receiver of the $u$-th D2D pair on the $k$-th resource block\\ \hline
	$g_{j,u,k}^{\rm D2D}$ & Channel gain between the transmitter of the $j$-th D2D pair and the receiver of the $u$-th D2D pair on the $k$-th resource block\\ \hline
	$I_{u,n,k}^{\rm D2M}$ & Interference from the transmitter of the $u$-th D2D pair to the $n$-th MUE on the $k$-th resource block\\ \hline
	$I_{j,u,k}^{\rm D2D}$ & Interference from the transmitter of the $j$-th D2D pair to the receiver of the $u$-th D2D pair on the $k$-th resource block\\ \hline
	$I_{n,u,k}^{\rm M2D}$ & Interference from the $n$-th MUE to the receiver of the $u$-th D2D pair on the $k$-th resource block\\ \hline
	$I_{n,k}^{\text{max}}$ & Maximum tolerable interference for the $n$-th MUE \\ \hline
	$\Phi$ & Set of unallocated resource blocks in the system \\ \hline
	$\Omega_u$ & Set of D2D users eligible for resource allocation \\ \hline
	$\Phi_u$ & Set of unallocated resource blocks for the $u$-th D2D user \\ \hline
	$\gamma_{u,k}^{\rm D}$ & SINR of the $u$-th D2D user on the $k$-th resource block \\ \hline
	$\gamma_{\rm th}$ & SINR threshold \\ \hline
	$I_{u,k}^{\rm D}$ & Interference caused by the $u$-th D2D user on the $k$-th resource block to the MUE \\ \hline
	$C_{u,k}^{\rm D}$ & Achieved capacity of the $u$-th D2D pair on the $k$-th resource block \\ \hline	
	$C_u^{\rm D}$ & Capacity of the $u$-th D2D pair \\ \hline
	$C_n^{\rm M}$ & Capacity of the $n$-th MUE \\ \hline
	$\Omega_k$ & Set of D2D users allocated to conflicting or invalid resource block $k$ \\ \hline
	$x$ & Each member of $\Omega_k$ \\ \hline
	$\alpha_{x,k}$ & Allocation status between resource block $k$ and member $x$ of $\Omega_k$ \\ \hline
	\end{tabular}
\end{table}

\subsection{Channel Model and Resource Allocation}
The channel model incorporates both large-scale fading (path loss, shadowing) and small-scale fading (multipath, Doppler spread)~\cite{9415745,7878038,7510820}.
Let $\alpha_{u,k}$ and $\alpha_{n,k}$ denote the binary resource allocation indicators for the $u$-th D2D pair and the $n$-th MUE on the $k$-th RB, respectively. They are defined as:
\begin{align}\label{eq:allocation_indicators}
\alpha_{u,k}, \alpha_{n,k} = 
\begin{cases}
1, & \text{if the } k\text{-th RB is allocated to the user,} \\
0, & \text{otherwise.}
\end{cases}
\end{align}
Similarly, we define the association indicators $\delta_{n,\rm B}, \delta_{n,v} \in \{0,1\}$ to represent the connection of the $n$-th MUE to the MBS or the $v$-th UAV. These must satisfy the constraint:
\begin{equation}\label{eq:mue_association_constraint}
\delta_{n,\rm B} + \sum_{v=1}^{N_{\rm V}} \delta_{n,v} \leq 1, \quad \forall n \in \{1,2,\ldots,N_{\rm M}\},
\end{equation}
where $N_{\rm M}$ is the total number of MUEs and $N_{\rm V}$ is the total number of UAVs.

\color{black}

When the $u$-th D2D pair uses the $k$-th resource block, let $P_{u,k}^{\rm D}$ denote the transmit power consumption of the $u$-th D2D pair on the $k$-th resource block, and $N_0$ denote the system noise. The signal-to-interference-plus-noise ratio (SINR) of the $u$-th D2D pair is given by:
\begin{align}\label{eq:SINR-D2D}
\gamma_{u,k}^{\rm D} = \dfrac {
\alpha_{u,k} P_{u,k}^{\rm D} g_{u,k}^{\rm D2D}
} {
\sum_{j=1,j\neq u}^{N_{\rm D}} I_{j,u,k}^{\rm D2D}
+ \sum_{n=1}^{N_{\rm M}} I_{n,u,k}^{\rm M2D}
+ N_0
}
\end{align}
where $g_{u,k}^{\rm D2D}$ is the channel gain between the transmitter and receiver of the $u$-th D2D pair on the $k$-th resource block;
the interference term $I_{j,u,k}^{\rm D2D} = \alpha_{j,k} P_{j,k}^{\rm D} g_{j,u,k}^{\rm D2D}$ is the interference from the transmitter of the $j$-th D2D pair to the receiver of the $u$-th D2D pair on the $k$-th resource block, and $g_{j,u,k}^{\rm D2D}$ is the channel gain between the transmitter of the $j$-th D2D pair and the receiver of the $u$-th D2D pair on the $k$-th resource block;
$I_{n,u,k}^{\rm M2D} = \alpha_{n,k} P_{n,k}^{\rm M} g_{n,u,k}^{\rm M2D}$ is the interference from the $n$-th MUE to the receiver of the $u$-th D2D pair on the $k$-th resource block, and $g_{n,u,k}^{\rm M2D}$ is the channel gain between the $n$-th MUE and the receiver of the $u$-th D2D pair on the $k$-th resource block.

When the $n$-th MUE uses the $k$-th resource block, let $P_{n,k}^{\rm M}$ denote the transmit power consumption of the $n$-th MUE on the $k$-th resource block. The signal-to-interference-plus-noise ratio (SINR) of the $n$-th MUE, $\gamma_{n,k}^{\rm M}$, is given by:
\begin{align}\label{eq:SINR-MUE}
	\gamma_{n,k}^{\rm M} = &\dfrac {
		\alpha_{n,k} P_{n,k}^{\rm M} \left(
		\delta_{n,\rm B}  g_{n,k}^{\rm M2B} +
		\sum_{v=1}^{N_{\rm V}}\delta_{n,v} g_{n,v,k}^{\rm M2V}
		\right)
	}{
		\sum_{u=1}^{N_{\rm D}} I_{u,n,k}^{\rm D2B}
		+ N_0
	}
\end{align}
where $g_{n,k}^{\rm M2B}$ is the channel gain between the $n$-th MUE and the MBS on the $k$-th resource block, and $g_{n,v,k}^{\rm M2V}$ is the channel gain between the $n$-th MUE and the $v$-th UAV on the $k$-th resource block. In the denominator of~\eqref{eq:SINR-MUE}, the first term is the total interference from all the D2D pairs to the serving base station of the $n$-th MUE on the $k$-th resource block. This is denoted as    
\begin{align}\label{eq:interference_from_D2D_to_MUE}
	I_{u,n,k}^{\rm D2B} =
		\delta_{n,\rm B} \alpha_{u,k} P_{u,k}^{\rm D} g_{u,k}^{\rm D2B} + 
		\sum_{v=1}^{N_{\rm V}} \delta_{n,v} \alpha_{u,k} P_{u,k}^{\rm D} g_{u,v,k}^{\rm D2V},
\end{align}
where $g_{u,k}^{\rm D2B}$ is the channel gain between the transmitter of the $u$-th D2D pair and the MBS on the $k$-th resource block, and $g_{u,v,k}^{\rm D2V}$ is the channel gain between the transmitter of the $u$-th D2D pair and the $v$-th UAV on the $k$-th resource block.

Let $W_{\rm RB}$ denote the bandwidth of each resource block. Based on the derived SINR expressions and Shannon's capacity formula, the total achievable data rates for the $u$-th D2D pair, denoted by $C_u^{\rm D}$, and the $n$-th MUE, denoted by $C_n^{\rm M}$, are computed by summing the capacity over all $K$ RBs:
\begin{align}
	\label{eq:user_capacity}
	C_u^{\rm D} &= \sum_{k=1}^K W_{\rm RB} \log_2\left(1+\gamma_{u,k}^{\rm D}\right), \nonumber \\
	C_n^{\rm M} &= \sum_{k=1}^K W_{\rm RB} \log_2\left(1+\gamma_{n,k}^{\rm M}\right).
\end{align}
Consequently, the total system capacity, denoted by $C_{\rm Total}$, is defined as the aggregate sum of the transmission capacities of all $N_{\rm D}$ D2D pairs and $N_{\rm M}$ MUEs:
\begin{align}\label{eq:total_capacity}
	C_{\rm Total} = \sum_{u=1}^{N_{\rm D}} C_u^{\rm D} + \sum_{n=1}^{N_{\rm M}} C_n^{\rm M}.
\end{align}

\color{black}
According to~\cite{7794815}, the power consumption of a terminal device consists of two parts: static circuit power consumption and transmission power consumption. Let $P_u^{\rm SC}$ denote the static circuit power consumption of the $u$-th D2D user pair, and $P_n^{\rm SC}$ denote the static circuit power consumption of the $n$-th MUE user. The reciprocal of the power amplifier efficiency is denoted by $\xi$.
Based on this, the power consumption of the $u$-th D2D user pair is 
\begin{align}\label{eq:total_power_of_a_D2D}
	P_u^{\rm D} = \xi \sum_{k=1}^K \alpha_{u,k} P_{u,k}^{\rm D} + P_u^{\rm SC}
\end{align}
and the power consumption of the $n$-th MUE user is
\begin{align}\label{eq:total_power_of_a_MUE}
	P_n^{\rm M} = \xi \sum_{k=1}^K \alpha_{n,k} P_{n,k}^{\rm M} + P_n^{\rm SC}.
\end{align}
According to~\eqref{eq:total_power_of_a_D2D}, the total power consumption of all the D2D users can be expressed as
\begin{align}\label{eq:total_power_of_all_D2D}
	P^{\rm D} = \sum_{u=1}^{N_{\rm D}} P_u^{\rm D}. 
\end{align}
Similarly, according to~\eqref{eq:total_power_of_a_MUE}, the total power consumption of all the MUEs can be expressed as
\begin{align}\label{eq:total_power_of_all_MUE}
	P^{\rm M} = \sum_{n=1}^{N_{\rm M}} P_n^{\rm M}. 
\end{align}

According to the above assumptions,  the energy efficiency of all D2D users in the system can be derived by
\begin{align}\label{eq:EE_of_all_D2D}
	EE^{\rm D} = \frac{C^{\rm D}}{P^{\rm D}} = \frac{\sum_{u=1}^{N_{\rm D}} C_u^{\rm D}}{\sum_{u=1}^{N_{\rm D}} P_u^{\rm D}}. 
\end{align}

We also consider the Jain's fairness index~\cite{R19} for the D2D users, which is a well-known metric for measuring fairness in resource allocation~\cite{7878038}. The Jain's fairness index is defined as the ratio of the square of the sum of the individual utilities to the sum of the squares of the individual utilities. It ranges from 0 to 1, where 0 indicates maximum unfairness and 1 indicates perfect fairness. In this case, we consider the transmission capacity of D2D users as their utility. Therefore, the Jain's fairness index for D2D users can be expressed as follows:

\begin{align}\label{eq:fairness_index}
	\mathcal{J}(\mathbf{R}) = \dfrac{\left(\sum_{u=1}^{N_{\rm D}}C_u^{\rm D}\right)^2}{N_{\rm D}\sum_{u=1}^{N_{\rm D}} \left(C_u^{\rm D}\right)^2}.
\end{align}

\section{Problem Formulation} 
\label{sec:problem_formulation}

Based on the system model described above, our primary objective is to maximize the total energy efficiency of D2D users $EE^{\mathrm{D}}$. However, simply maximizing sum-EE without fairness constraints typically results in resource starvation for cell-edge users with poor channel conditions. To prevent this, we explicitly incorporate a fairness constraint into the optimization framework. The joint optimization problem is formulated as follows:
\begin{subequations}
	\begin{align}
    (\text{P1}): \quad & \max_{\alpha_{u,k}, P_{u,k}^{\mathrm{D}}} \quad EE^{\mathrm{D}} \label{eq:obj} \\
    \text{s.t.} \quad 
    & \alpha_{u,k} \in \{0,1\}, \quad \forall u, \forall k, \label{eq:c1} \\
    & \alpha_{u,k} \gamma_{u,k}^{\mathrm{D}} \geq \gamma_{\mathrm{th}}, \text{ if } \alpha_{u,k} = 1, \quad \forall u, \forall k,  \label{eq:c2} \\
    & P_{u,k}^{\mathrm{D}} \leq P_{\mathrm{max}}, \quad \forall u, \forall k, \label{eq:c3} \\
    & \sum_{u=1}^{N_{\mathrm{D}}} \alpha_{u,k} I_{u,n,k}^{\mathrm{D2M}} \leq I_{n,k}^{\mathrm{max}}, \quad \forall k, \forall n, \label{eq:c4} \\
    & P_n^{\mathrm{M}} \leq P_{\mathrm{max}}, \quad \forall n, \label{eq:c5} \\
    & \mathcal{J}(\mathbf{R}) \geq F_{\mathrm{th}}. \label{eq:c6}
\end{align}
\end{subequations}

Constraint \eqref{eq:c1} ensures binary resource allocation. Constraint \eqref{eq:c2} guarantees the minimum QoS requirement (SINR threshold $\gamma_{\mathrm{th}}$). Constraints \eqref{eq:c3} and \eqref{eq:c5} limit the maximum transmit power for D2D users and MUEs, respectively. Constraint \eqref{eq:c4} manages the aggregate interference to MUEs. Most importantly, Constraint \eqref{eq:c6} enforces fairness, requiring that the D2D users' fairness index $\mathcal{J}(\mathbf{R})$ remains above a predefined threshold $F_{\mathrm{th}} \in [0, 1]$.

\vspace{0.5em} 
\noindent \textbf{Motivation for Heuristic Approach:} 
The formulation of (P1) presents a formidable computational challenge due to two main factors. 
First, unlike exclusive reuse schemes where interference is deterministic, our system supports \textit{shared reuse} (as highlighted in Table I), allowing multiple D2D pairs to multiplex on the same RB. This introduces complex \textit{many-to-many interference couplings} in the denominator of the SINR term, making the feasible region highly non-convex.
Second, the fairness index in Constraint \eqref{eq:c6} is a fractional and non-convex function of the data rates, which conflicts with the sum-EE objective.

Consequently, (P1) is a non-convex Mixed-Integer Non-Linear Programming (MINLP) problem, which is known to be NP-hard. Obtaining a global optimal solution would require exponential complexity, which is intractable for real-time UAV-assisted networks. This mathematical intractability necessitates an efficient low-complexity solution. Therefore, instead of attempting to solve (P1) via prohibitive exhaustive search, we propose the \textit{Multi-Channel Rate-Fair} (MCRF) algorithm. The MCRF adopts a greedy heuristic strategy that satisfies the fairness intention of Constraint \eqref{eq:c6} by dynamically prioritizing users with lower historical rates, thereby achieving a high-quality sub-optimal solution suitable for dynamic AGIHN environments.

\section{The Proposed Multi-Channel Rate Fair (MCRF) Allocation} 
\label{sec:proposed_method}
In this section, we first introduce the problem transformation and the overall framework of the proposed MCRF method. Next, we present several preprocessed channel gain expressions among D2D users and MUEs to facilitate understanding of the detailed algorithm in the following subsections. After that, we elaborate on the re-association schemes incorporated within the MCRF framework. Subsequently, we describe the detailed procedure of the MCRF method. Finally, we discuss the advantages and benefits of the proposed design.

\subsection{Problem Transformation and Framework of MCRF}

To simplify the mathematically intractable joint optimization problem (P1), we adopt a heuristic approximation by decoupling resource allocation from dynamic power control. Specifically, we assign a fixed equal transmission power $P_{u,k}^{\mathrm{D}} = \frac{P_{\mathrm{max}}}{K}$ for each D2D user across available channels. This assumption evenly distributes power, reducing signaling overhead and computational complexity. 

Consequently, the optimization focus shifts entirely to determining the subchannel assignment indicator $\alpha_{u,k}$. The simplified resource assignment problem (P2) solved by the proposed MCRF algorithm is formulated as follows:
\begin{subequations} \label{eq:p2}
\begin{align}
    (\text{P2}): \quad & \max_{\alpha_{u,k}} \quad EE^{\mathrm{D}} \label{eq:p2_obj} \\
    \text{s.t.} \quad 
    & \text{Constraints } \eqref{eq:c1}, \eqref{eq:c2}, \eqref{eq:c4}, \eqref{eq:c5}, \eqref{eq:c6}, \notag \\
    & P_{u,k}^{\mathrm{D}} = \frac{P_{\mathrm{max}}}{K}, \quad \forall u, \forall k. \label{eq:p2_c3}
\end{align}
\end{subequations}

\begin{figure}[!t]
	\centering
	\includegraphics[width=\columnwidth]{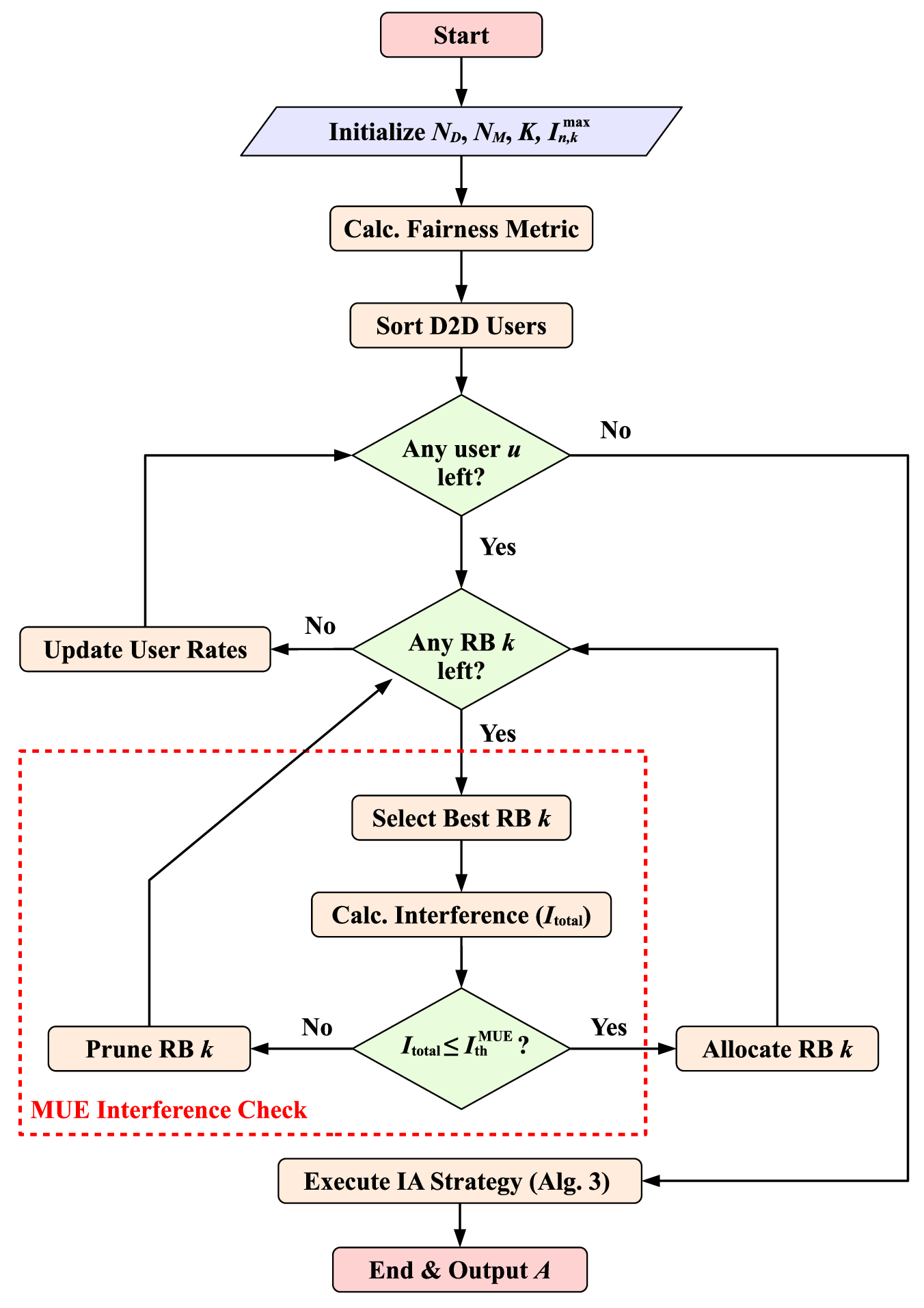}
	\caption{Flowchart of the proposed MCRF algorithm.}
	\label{fig:flowchart}
\end{figure}

By replacing the inequality power constraint with the equality constraint in \eqref{eq:p2_c3}, the computationally prohibitive joint optimization is transformed into a tractable integer programming problem. To protect the MUEs from aggregate interference without relying on dynamic power tuning, MCRF relies on the macro base station and UAV base stations to broadcast a maximum interference tolerance threshold ($I_{n,k}^{\mathrm{max}}$). The algorithm incorporates a strict cross-tier interference constraint check (highlighted in the dashed box of Fig.~\ref{fig:flowchart}), which acts as a spectral gatekeeper. For every RB candidate, it calculates the potential aggregate interference to the associated MUE. If allocating the RB to a D2D pair violates this strict tolerance threshold, the candidate RB is immediately pruned. 

Furthermore, because the proposed framework supports shared spectrum reuse, multiple D2D pairs may multiplex on the same RB, leading to severe intra-tier interference. To resolve this, the framework executes an independent IA strategy as a post-allocation refinement. This IA strategy dynamically revokes the allocation of D2D users experiencing poor SINR to strictly preserve the transmission quality and fairness among the remaining D2D users sharing that RB.

While this equal-power approximation incurs a slight loss in theoretical peak throughput compared to ideal joint optimization, it guarantees robust MUE protection, effectively mitigates D2D intra-tier interference, ensures strict user fairness, and maintains a low polynomial complexity suitable for real-time execution in dynamic AGIHN scenarios.

\color{black}

The rate-fair allocation algorithm ensures the quality of service (QoS) for MUEs while maintaining fairness among D2D users. It is applicable in different scenarios: single-channel and multi-channel. In the single-channel scenario, which serves as a baseline and a special case, a single resource block can only be reused once by one D2D user. In the multi-channel scenario, different D2D users can reuse a single resource block multiple times to enhance system spectral efficiency and increase system capacity. Additionally, it incorporates an interference avoidance strategy to address interference issues among D2D users.

\subsection{Preprocessing}
To improve the efficiency of interference and link calculations in subsequent resource allocation algorithms, this subsection first defines several preprocessed expressions for channel gains.

First, to simplify the calculation of interference from D2D pairs to MUEs, we predefine $g_{u,n,k}^{\rm D2M}$, which represents the interference channel gain from the $u$-th D2D pair to the $n$-th MUE when transmitting on resource block $k$. It is calculated as follows:
\begin{align}\label{eq:D2M_gain}
g_{u,n,k}^{\rm D2M} = \delta_{n,\rm B}g_{u,k}^{\rm D2B} + \sum_{v=1}^{N_{\rm V}} \delta_{n,v} g_{u,v,k}^{\rm D2V}.
\end{align}


Next, to facilitate the calculation of the signal transmission path gain for MUEs, we define $g_{n,k}^{\rm M2B}$ as the effective channel gain between the $n$-th MUE and its associated base station (MBS or UAV) on resource block $k$, given by
\begin{align}\label{eq:M2B_gain}
g_{n,k}^{\rm M2B} = \delta_{n,\rm B} g_{n,k}^{\rm M2B} + \sum_{v=1}^{N_{\rm V}} \delta_{n,v} g_{n,v,k}^{\rm M2V},
\end{align}
where $g_{n,k}^{\rm M2B}$ is the channel gain between the $n$-th MUE and the MBS on resource block $k$, and $g_{n,v,k}^{\rm M2V}$ is the channel gain between the $n$-th MUE and the $v$-th UAV on resource block $k$.


This study focuses on the resource allocation strategy for D2D users. Therefore, we set $N_{\rm M} = K$ to simplify the system model and avoid considering interference among MUEs. 
First, each MUE decides whether to connect to a UAV or the MBS. The selection criterion is to maximize the average channel gain, i.e., $\max(\sum_{k=1}^{K} g_{n,k}^{\rm {M2B}})$. Next, for each resource block $k$, one MUE $n$ is selected, $\alpha_{n,k}$ is set to $1$, and the $n$-th MUE is removed from the set of candidates. In this way, each MUE is assigned a dedicated resource block, and there is no mutual interference among MUEs.

\subsection{Single-Channel Case}

The single-channel rate fair resource allocation strategy can be divided into two steps:
\begin{itemize}
	\item Step 1: Each D2D pair sequentially selects one resource block ($k^*$). The selection criterion is to maximize the ratio of the D2D signal transmission channel gain to the total interference channel gain to all MUEs using the $k^*$-th resource block. Specifically, for the $u$-th D2D pair, the selection is given by
    \begin{align}
        \label{eq:step1_selection}
        k^* =& \arg\max_{k \in \Phi_u} \frac{g_{u,k}^{\mathrm{D2D}}}{\sum_{n=1}^{N_{\mathrm{M}}} \alpha_{n,k}I_{u,n,k}^{\mathrm{D2M}}}\notag \\ 
        \text{ s.t. }& \sum_{n=1}^{N_{\mathrm{M}}} \alpha_{n,k} I_{u,n,k}^{\mathrm{D2M}} \le \sum_{n=1}^{N_{\mathrm{M}}} \alpha_{n,k} I_{n,k}^{\mathrm{max}},
    \end{align}	
	\color{black}
    where $\Phi_u$ is the set of available resource blocks for the $u$-th D2D pair, $\alpha_{n,k}$ is the allocation status of the $n$-th MUE on the $k$-th RB, $I_{u,n,k}^{\mathrm{D2M}}$ is the interference caused by the $u$-th D2D pair on the $k$-th resource block to the $n$-th MUE, and $I_{n,k}^{\mathrm{max}}$ is the maximum tolerable interference for the $n$-th MUE, which can be calculated as
    \begin{align}\label{eq:max_interference}
        I_{n,k}^{\mathrm{max}} = \frac{P_{n,k}^{\mathrm{M}} g_{n,k}^{\mathrm{M2B}}}{\gamma_{\mathrm{th}}}.
    \end{align}
    This process continues until all users have completed one round of selection.

	\item Step 2: Select the D2D pair ($u^*$) with the lowest current system transmission capacity and allow this pair to prioritize selecting a resource block ($k^*$). The selection criterion remains:
	\begin{align}
		\label{eq:step2_selection}
		k^* =& \arg\max_{k \in \Phi_{u^*}} \frac{g_{u^*,k}^{\mathrm{D2D}}}{\sum_{n=1}^{N_{\mathrm{M}}} \alpha_{n,k} I_{u^*,n,k}^{\mathrm{D2M}}} \notag \\ 
		\text{ s.t. }& \sum_{n=1}^{N_{\mathrm{M}}} \alpha_{n,k} I_{u^*,n,k}^{\mathrm{D2M}} \le \sum_{n=1}^{N_{\mathrm{M}}} \alpha_{n,k} I_{n,k}^{\mathrm{max}}, 
	\end{align}
	\color{black}
	where $\Phi_{u^*}$ is the set of available resource blocks for the $u^*$-th D2D pair. The selection criterion is the same as in Step 1. 
	Note that the operation at Line~\ref{alg:SCRF:line:24} in Algorithm~\ref{alg:SCRF} removes the selected resource block from all sets of available resource blocks ($\Phi_{u}$) for all D2D pairs ($\forall u$), ensuring that the selected RB is not simultaneously assigned to multiple D2D pairs. Step 2 continues until no further resource blocks can be allocated.
\end{itemize}

These two steps ensure that the selected resource block maintains the transmission quality of the MUE. In the first step, each D2D pair selects a resource block, while in the second step, priority is given to the D2D pair with the lowest system transmission capacity. This approach effectively enhances the overall system performance in the single-channel scenario. The detailed single-channel rate-fair algorithm is shown in Algorithm~\ref{alg:SCRF}.

\begin{algorithm2e}[!t]
\caption{Single-Channel Rate-Fair (SCRF) Algorithm}
\label{alg:SCRF}
\SetAlgoLined
\KwIn{Number of D2D users $N_D$, Number of cellular users $N_M$, Number of RBs $K$, CSI, Threshold $I_{n,k}^{\mathrm{max}}$}
\KwOut{Allocation Matrix $\mathbf{A} = [\alpha_{u,k}]$}

\textbf{Initialize:} Unallocated users $\Omega_U \leftarrow \{1, \dots, N_D\}$, Available RBs $\Phi_u \leftarrow \{1, \dots, K\}$ for all $u$, $\alpha_{u,k} \leftarrow 0$, $I_{n,k}^{\mathrm{cur}} \leftarrow 0$, $\forall u, k$\;

\tcp{Step 1: Sequential Selection}
\ForEach{user $u \in \{1, \dots, N_D\}$}{
    $k^* \leftarrow \arg \max_{k \in \Phi_u} \left( \frac{g_{u,k}^{\mathrm{D2D}}}{\sum_{n=1}^{N_M} \alpha_{n,k} g_{u,n,k}^{\mathrm{D2M}}} \right)$\;
    Calculate interference: $I_{\mathrm{temp}} \leftarrow I_{n,k^*}^{\mathrm{cur}} + P_{u,k^*}^{\mathrm{D}} g_{u,n,k^*}^{\mathrm{D2M}}$\;
    
    \While{$I_{\mathrm{temp}} > I_{n,k^*}^{\mathrm{max}}$ \textbf{and} $\Phi_u \neq \emptyset$}{
        $\Phi_u \leftarrow \Phi_u \setminus \{k^*\}$\;
        Re-select $k^*$ from $\Phi_u$ using max ratio criteria\;
    }
    
    \If{valid $k^*$ found}{
        $\alpha_{u,k^*} \leftarrow 1$; $I_{n,k^*}^{\mathrm{cur}} \leftarrow I_{\mathrm{temp}}$\;
        \tcp{Global Removal (Exclusive Access)}
        $\Phi_j \leftarrow \Phi_j \setminus \{k^*\}, \forall j \in \{1, \dots, N_D\}$\;
    }
    \If{$\Phi_u = \emptyset$}{ $\Omega_U \leftarrow \Omega_U \setminus \{u\}$\; }
}

\tcp{Step 2: Fairness-Aware Re-allocation}
Calculate $\gamma_{u,k}^{\mathrm{D}}$ and $C_{u,k}^{\mathrm{D}}$ for all $u, k$\;
\While{$\Omega_U \neq \emptyset$}{
    $u^* \leftarrow \arg \min_{u \in \Omega_U} \sum_{k=1}^K \alpha_{u,k} C_{u,k}^{\mathrm{D}}$\;
    Find best RB $k^*$ for $u^*$ in $\Phi_{u^*}$ satisfying constraints\;
    
    \eIf{valid $k^*$ found}{
        $\alpha_{u^*,k^*} \leftarrow 1$; Update $I_{n,k^*}^{\mathrm{cur}}$\;
        $\Phi_j \leftarrow \Phi_j \setminus \{k^*\}, \forall j \in \{1, \dots, N_D\}$\label{alg:SCRF:line:24}\;
    }{
        $\Omega_U \leftarrow \Omega_U \setminus \{u^*\}$\;
    }
    Update $\gamma_{u,k}^{\mathrm{D}}, C_{u,k}^{\mathrm{D}}$\;
}
\Return $\mathbf{A}$
\end{algorithm2e}

\subsection{Multi-Channel Case}
In the multi-channel case, the resource allocation strategy is similar to the single-channel case. However, in this case, multiple D2D users can reuse a single resource block. The multi-channel rate fair resource allocation can be divided into two steps:
\begin{itemize}
    \item Step 1: Each D2D user sequentially selects one resource block ($k^*$). The selection criterion is based on the maximum ratio of the D2D signal transmission channel gain to the total interference channel gain to all MUEs using the $k^*$-th resource block. Specifically, for the $u$-th D2D pair, the selection is given by
    \begin{align}
        \label{eq:30}
        k^* =& \arg\max_{k \in \Phi_u} \frac{g_{u,k}^{\mathrm{D2D}}}{\sum_{n=1}^{N_{\mathrm{M}}} \alpha_{n,k} I_{u,n,k}^{\mathrm{D2M}}} \notag \\
        \text{ s.t. }& \sum_{n=1}^{N_{\mathrm{M}}} \alpha_{n,k} \left( I_{n,k}^{\mathrm{cur}} + I_{u,n,k}^{\mathrm{D2M}} \right) \le \sum_{n=1}^{N_{\mathrm{M}}} \alpha_{n,k} I_{n,k}^{\mathrm{max}},
    \end{align}
	\color{black}
    where $\Phi_u$ is the set of available resource blocks for the $u$-th D2D pair, $\alpha_{n,k}$ is the allocation status of the $n$-th MUE on the $k$-th RB, 
	$I_{n,k}^{\mathrm{cur}}$ is the current aggregate interference on the $k$-th RB to the $n$-th MUE from already allocated D2D pairs, 
	\color{black}
	$I_{u,n,k}^{\mathrm{D2M}}$ is the interference caused by the $u$-th D2D pair on the $k$-th resource block to the $n$-th MUE, and $I_{n,k}^{\mathrm{max}}$ is the maximum tolerable interference for the $n$-th MUE, which can be obtained by
    \begin{align}\label{eq:31}
        I_{n,k}^{\mathrm{max}} = \frac{P_{n,k}^{\mathrm{M}} g_{n,k}^{\mathrm{M2B}}}{\gamma_{\mathrm{th}}}.
    \end{align}
    This process continues until all users have completed one round of selection.
    
    \item Step 2: Select the D2D user ($u^*$) with the lowest current system transmission capacity and allow this user to prioritize selecting a resource block ($k^*$). The selection criterion remains:
    \begin{align}\label{eq:32}
        k^* =& \arg\max_{k \in \Phi_{u^*}} \frac{g_{u^*,k}^{\mathrm{D2D}}}{\sum_{n=1}^{N_{\mathrm{M}}} \alpha_{n,k} I_{u^*,n,k}^{\mathrm{D2M}}} \notag \\
        \text{ s.t. }& \sum_{n=1}^{N_{\mathrm{M}}} \alpha_{n,k} \left( I_{n,k}^{\mathrm{cur}} + I_{u^*,n,k}^{\mathrm{D2M}} \right) \le \sum_{n=1}^{N_{\mathrm{M}}} \alpha_{n,k} I_{n,k}^{\mathrm{max}},
    \end{align}
	\color{black}
    where $\Phi_{u^*}$ is the set of available resource blocks for the $u^*$-th D2D pair. The selection criterion is the same as in Step 1. Note that the operation at Line~\ref{alg:MCRF:line:23} in Algorithm~\ref{alg:MCRF} is different from the single-channel case. In the multi-channel case, it only removes the selected resource block from the set of available resource blocks ($\Phi_{u^*}$) for the $u^*$-th D2D pair, allowing other D2D users to reuse this resource block. Step 2 continues until no further resource blocks can be allocated. 
\end{itemize}

The multi-channel rate fair resource allocation algorithm ensures that the selected resource block maintains the transmission quality of the MUE. In the first step, each pair of D2D users selects a resource block, while in the second step, priority is given to the D2D user with the lowest system transmission capacity. This approach effectively enhances the overall system performance in the multi-channel scenario. The detailed multi-channel rate-fair algorithm is shown in Algorithm~\ref{alg:MCRF}.

\begin{algorithm2e}[t]
\caption{Multi-Channel Rate-Fair (MCRF) Algorithm}
\label{alg:MCRF}
\SetAlgoLined
\KwIn{Number of D2D users $N_D$, Number of cellular users $N_M$, Number of RBs $K$, CSI, Threshold $I_{n,k}^{\mathrm{max}}$}
\KwOut{Allocation Matrix $\mathbf{A} = [\alpha_{u,k}]$}

\textbf{Initialize:} $\Omega_U \leftarrow \{1, \dots, N_D\}$, $\Phi_u \leftarrow \{1, \dots, K\}$ for all $u$, $\alpha_{u,k} \leftarrow 0$, $I_{n,k}^{\mathrm{cur}} \leftarrow 0$, $\forall u, k$\;

\tcp{Step 1: Initial Sequential Selection}
\ForEach{user $u \in \{1, \dots, N_D\}$}{
    $k^* \leftarrow \arg \max_{k \in \Phi_u} \left( \frac{g_{u,k}^{\mathrm{D2D}}}{\sum_{n=1}^{N_M} \alpha_{n,k} g_{u,n,k}^{\mathrm{D2M}}} \right)$\;
    Calculate interference: $I_{\mathrm{temp}} \leftarrow I_{n,k^*}^{\mathrm{cur}} + P_{u,k^*}^{\mathrm{D}} g_{u,n,k^*}^{\mathrm{D2M}}$\;
    
    \While{$I_{\mathrm{temp}} > I_{n,k^*}^{\mathrm{max}}$ \textbf{and} $\Phi_u \neq \emptyset$}{
        $\Phi_u \leftarrow \Phi_u \setminus \{k^*\}$; Re-select $k^*$\;
    }
    \If{valid $k^*$ found}{
        $\alpha_{u,k^*} \leftarrow 1$; $I_{n,k^*}^{\mathrm{cur}} \leftarrow I_{\mathrm{temp}}$\;
        \tcp{Local Removal (Allow Reuse)}
        $\Phi_u \leftarrow \Phi_u \setminus \{k^*\}$\;
    }
    \If{$\Phi_u = \emptyset$}{ $\Omega_U \leftarrow \Omega_U \setminus \{u\}$\; }
}

\tcp{Step 2: Iterative Fairness Enhancement}
Calculate $\gamma_{u,k}^{\mathrm{D}}$ and $C_{u,k}^{\mathrm{D}}$\;
\While{$\Omega_U \neq \emptyset$}{
    $u^* \leftarrow \arg \min_{u \in \Omega_U} \sum_{k=1}^K \alpha_{u,k} C_{u,k}^{\mathrm{D}}$\;
    Find best RB $k^*$ for $u^*$ in $\Phi_{u^*}$ satisfying constraints\;
    
    \eIf{valid $k^*$ found}{
        $\alpha_{u^*,k^*} \leftarrow 1$; Update $I_{n,k^*}^{\mathrm{cur}}$\;
        $\Phi_{u^*} \leftarrow \Phi_{u^*} \setminus \{k^*\}$; Update $C_{u^*}^D$\label{alg:MCRF:line:23}\;
    }{
        $\Omega_U \leftarrow \Omega_U \setminus \{u^*\}$\;
    }
}
\Return $\mathbf{A}$
\end{algorithm2e}

\subsection{Interference Avoidance Strategy}
The interference avoidance strategy is designed to prevent interference among D2D users. The principle of this strategy is as follows: if the SINR of a D2D pair users on a specific resource block is below the SINR threshold, the allocation status of that user pair on the resource block will be revoked. This process continues until all pairs of D2D users allocated to a resource block meet the SINR threshold or are removed.

The detailed interference avoidance strategy can be described as follows:
\begin{itemize}
	\item Step 1: For each resource block $k$, check the allocation status of each D2D pair $u$. If the allocation status $\alpha_{u,k} = 1$ and the SINR $\gamma_{u,k}^{\rm D} < \gamma_{\rm th}$, add the D2D pair $u$ to the set of unqualified users $\Omega_k^{\rm U}$ for resource block $k$.
	\item Step 2: While the set of unqualified users $\Omega_k^{\rm U}$ is not empty, check the number of unqualified users:
		\begin{itemize}
			\item If there is only one unqualified user $u^*$ in $\Omega_k^{\rm U}$, revoke the allocation status of this user on resource block $k$ by setting $\alpha_{u^*,k} = 0$. Update the SINR and capacity for all D2D pairs, and remove $u^*$ from $\Omega_k^{\rm U}$.
			\item If there are multiple unqualified users in $\Omega_k^{\rm U}$, select the user $u^*$ with the maximum ratio of total capacity to SINR, revoke its allocation status on resource block $k$, update the SINR and capacity for all D2D pairs, and remove $u^*$ from $\Omega_k^{\rm U}$. Then, check the remaining unqualified users in $\Omega_k^{\rm U}$ and remove those that now meet the SINR threshold.
		\end{itemize}
	Repeat this process until the set of unqualified users $\Omega_k^{\rm U}$ is empty.
\end{itemize}

The detailed process of interference avoidance is shown in Algorithm~\ref{alg:IA}.

\begin{algorithm2e}[!t]
\caption{Interference Avoidance (IA) Strategy}
\label{alg:IA}
\SetAlgoLined
\KwIn{Initial Allocation $\mathbf{A}$, SINR Threshold $\gamma_{\mathrm{th}}$}
\KwOut{Refined Allocation $\mathbf{A} = [\alpha_{u,k}]$}

\ForEach{RB $k \in \{1, \dots, K\}$}{
    \tcp{Identify users violating SINR threshold}
    $\Omega_k^{\mathrm{U}} \leftarrow \{u \mid \alpha_{u,k}=1 \textbf{ and } \gamma_{u,k}^{\mathrm{D}} < \gamma_{\mathrm{th}} \}$\;
    
    \While{$\Omega_k^{\mathrm{U}} \neq \emptyset$}{
        \eIf{$|\Omega_k^{\mathrm{U}}| == 1$}{
            $u^* \leftarrow$ the single element in $\Omega_k^{\mathrm{U}}$\;
        }{
            \tcp{Remove user with highest Capacity-to-SINR ratio}
            $u^* \leftarrow \arg \max_{u \in \Omega_k^{\mathrm{U}}} \left( \frac{\sum_{k'} \alpha_{u,k'} C_{u,k'}^{\mathrm{D}}}{\gamma_{u,k}^{\mathrm{D}}} \right)$\;
        }
        
        $\alpha_{u^*,k} \leftarrow 0$; Remove $u^*$ from $\Omega_k^{\mathrm{U}}$\;
        Update SINR and Capacity for remaining users on RB $k$\;
        
        \tcp{Re-check qualification of remaining users}
        \ForEach{$u \in \Omega_k^{\mathrm{U}}$}{
            \If{$\alpha_{u,k} \gamma_{u,k}^{\mathrm{D}} \ge \gamma_{\mathrm{th}}$}{
                $\Omega_k^{\mathrm{U}} \leftarrow \Omega_k^{\mathrm{U}} \setminus \{u\}$ \tcp*[r]{User recovered}
            }
        }
    }
}
\Return $\mathbf{A}$
\end{algorithm2e}

\section{Complexity Analysis}
\label{sec:complexity}

In this section, we analyze the computational complexity of the proposed algorithm and provide a qualitative comparison with emerging learning-based strategies. The goal is to demonstrate that the proposed heuristic strikes a favorable balance between performance and computational efficiency, particularly for dynamic UAV-assisted environments.

\subsection{Computational Complexity and Runtime Analysis}

The computational cost of the proposed MCRF scheme is dominated by two sequential phases: the iterative resource allocation and the Interference Avoidance (IA) strategy.

First, in the resource allocation phase, the algorithm iteratively selects the most suitable RB for each D2D pair. For a system with $N_D$ D2D pairs and $K$ resource blocks, constructing the preference list involves calculating the interference leakage to $N_M$ MUEs, which incurs a complexity of $O(N_M K)$. Since the algorithm dynamically prioritizes users based on historical rates and may update candidate lists in each iteration, the worst-case complexity for allocating resources to all $N_D$ users is dominated by $O(N_D N_M K + N_D^2 K)$.

Second, the IA phase verifies the SINR requirements for each RB. In the worst-case scenario, where multiple users share the same RB, the strategy must check the SINR conditions and potentially iteratively remove the user with the maximum capacity-to-SINR ratio. Checking and updating the status for $N_D$ users across $K$ resource blocks requires $O(N_D^2 K)$ operations.

Consequently, the overall computational complexity of the proposed framework is the sum of these two phases, which can be expressed as:
\begin{equation}
    \mathcal{O}_{\text{total}} = O(N_{D}N_{M}K + N_{D}^{2}K).
\end{equation}
Since $N_D$, $N_M$, and $K$ are finite parameters, the proposed algorithm exhibits polynomial time complexity. This indicates that the solution is highly scalable and feasible for implementation on standard processors without requiring computationally intensive hardware.

\vspace{.5em}
\noindent \textbf{Actual Runtime Evaluation and Practical Comparison:} 
To validate the real-time feasibility of the proposed algorithm, actual execution times were measured using a standard hardware workstation (AMD Ryzen 9 5950X CPU, 64 GB RAM) running MATLAB, without the benefit of hardware pipelining or parallel processing optimizations. For a representative network size of $N_D = 50$ D2D pairs and $K = 50$ subchannels, the average execution time per allocation snapshot is $258.48$ ms. For a denser network with $N_D = 100$ and $K = 50$, the runtime scales to $769.02$ ms, which is consistent with the theoretically derived polynomial complexity bound.

In practical cellular systems (e.g., LTE and 5G NR), a typical baseband scheduling slot (i.e., Transmission Time Interval) is strictly bounded to $1$ ms. Although the measured software-based runtime exceeds this interval, it is crucial to recognize the fundamental distinction between software-level simulation and practical hardware implementation. Real-world resource allocation algorithms are not executed on general-purpose CPUs using interpreted languages. Instead, they are synthesized onto dedicated hardware accelerators, such as \textit{Field-Programmable Gate Arrays} (FPGAs) or \textit{Application-Specific Integrated Circuits} (ASICs). Executing complex iterative algorithms on dedicated parallelized hardware reduces the processing latency by orders of magnitude to the microsecond ($\mu$s) scale. Given the proven low polynomial complexity of the proposed MCRF algorithm, it can comfortably operate within the standard $1$ ms scheduling slot, demonstrating its feasibility for real-time, frame-by-frame resource allocation in highly dynamic AGIHN environments.

\subsection{Comparison with Learning-based Approaches}
To justify the practicality of our heuristic approach, we contrast it with state-of-the-art Deep Reinforcement Learning (DRL) schemes (e.g., \cite{Noman2025}), as highlighted in Table \ref{tab:comparison}. While DRL methods can theoretically approximate optimal policies, they face critical limitations in dynamic AGIHNs:

\begin{itemize}
    \item \textbf{Training Overhead vs. Training-Free:} DRL agents require extensive offline training episodes and frequent online fine-tuning to converge to a stable policy. In disaster relief missions where UAVs are deployed ad-hoc, the time cost for training may exceed the mission's critical response window. In contrast, our MCRF algorithm is training-free and deterministic, allowing for immediate deployment.
    
    \item \textbf{Adaptability to Topology Changes:} The high mobility of UAVs causes rapid variations in channel gains and interference patterns. Learning-based models are often sensitive to such distribution shifts and may require retraining when the topology changes. Conversely, our optimization-based heuristic executes in real-time (in the order of milliseconds per time slot), ensuring that resource allocation always reflects the \textit{current} instantaneous CSI.
\end{itemize}

In summary, while learning-based methods offer a promising direction for static scenarios, the proposed MCRF algorithm provides a more robust and low-latency solution for time-sensitive and mobility-aware UAV networks.

\section{Simulation Results}
\label{sec:simulation}

\subsection{Simulation Environment}
The simulation parameters adopted in this study are summarized in Table~\ref{tab:simulation_parameters}, which are configured in accordance with the standards in~\cite{6364435,7794815,7878038}. We consider a single macrocell environment with a radius of 500 m. The system consists of $N_{\mathrm{M}} = 50$ MUEs, while the number of D2D pairs $N_{\mathrm{D}}$ varies from 10 to 100 to evaluate system performance under different densities. Unless otherwise stated, the default number of D2D pairs is set to 10.

Regarding the channel model, the path loss for MUE links (Cellular links) is calculated as $128.1 + 37.6 \log_{10}(d_{\mathrm{MUE}})$, where $d_{\mathrm{MUE}}$ is the distance in kilometers. For D2D links, the path loss model is given by $148 + 40 \log_{10}(d_{\mathrm{D2D}})$. Small-scale fading is modeled using a Rayleigh distribution.

The bandwidth of each Resource Block (RB) is 180 kHz, and the thermal noise power spectral density is set to -174 dBm/Hz. The maximum transmission power for each device is limited to 23 dBm. The static circuit power consumption ($P^{\mathrm{SC}}$) and the power amplifier efficiency ($\xi$) are set to 100 mW and 0.38, respectively. The SINR threshold for QoS requirements is set to 3 dB. All simulations are conducted using MATLAB, and the results are averaged over 1000 Monte Carlo iterations to ensure statistical accuracy.

\vspace{0.5em} 
\noindent \textbf{Remark on User Topology:} 
It is worth noting that while the proposed MCRF algorithm is designed to handle arbitrary user locations in real-world deployment, in our simulations, we deliberately set the distance between D2D pairs ($d_{\mathrm{D2D}}$) to fixed values ranging from 10 m to 100 m. This controlled setup serves as a sensitivity analysis to strictly evaluate how proximity affects system metrics such as fairness and throughput, without the variance introduced by random mobility models. The detailed impact of varying $d_{\mathrm{D2D}}$ is explicitly evaluated and discussed in Section~\ref{subsec:impact_d2d_distance}.

\begin{table}[!t]
    \centering
    \label{tab:simulation_parameters}
    \caption{Simulation Parameters}
    \begin{tabular}{|m{2.8cm}|c|l|}
        \hline
        \textbf{Parameter} & \textbf{Symbol} & \textbf{Value} \\ \hline
        Macrocell radius & $R_M$ & 500 m \\ \hline
        Number of MUs & $N_{\rm M}$ & 50 \\ \hline
        Number of D2D pairs & $N_{\rm D}$ & 10, 20, \dots, 100 \\ \hline
        Number of resource blocks & $k$ & 50 \\ \hline
        D2D distance & $d_{\rm D2D}$ & 10m, 20m, \dots, 100m\\ \hline
		Distance between MUE and BS & $d_{\rm MUE}$ & 1m, 2m, \dots, 500m \\ \hline
        RB bandwidth & $W_{\rm RB}$ & 180 kHz \\ \hline
        Power spectral density of noise & $N_0$ & -174 dBm/Hz \\ \hline
        Pathloss model for MUE links & $g_{n,k}^{\rm M2B}$ & $128.1 + 37.6 \log(d_{\rm MUE}~[\text{km}])$ \\ \hline
        Pathloss model for D2D links & $g_{u,k}^{\rm D2D}$ & $148 + 40 \log(d_{\rm D2D}~[\text{km}])$ \\ \hline
        Maximum power for Device& $P_{\rm max}$ & 23 dBm \\ \hline 
        Channel fading model & - & Rayleigh \\ \hline
        SINR threshold & $\gamma_{\rm th}$ & 3 dB \\ \hline
        Static circuit power of device & $P_u^{\rm SC}, P_n^{\rm SC}$ & 100 mW \\ \hline
        PA efficiency of each Device & $\xi$ & 0.38 \\ \hline
    \end{tabular}
\end{table}

\subsection{Comparative Methods}
Existing literature indicates a fundamental trade-off in resource allocation: the Max-SINR algorithm maximizes system capacity at the expense of fairness, whereas the Round Robin (RR) algorithm guarantees fairness but yields suboptimal throughput. In our simulations, we benchmark the proposed strategy against these conventional schemes under two distinct reuse modes:

\begin{itemize}
    \item \textbf{Single-Channel Round Robin (SCRR):} A fairness-centric baseline where D2D users take turns selecting the most suitable RB based on the signal-to-leakage ratio, ensuring minimal impact on MUEs.
    \item \textbf{Single-Channel Max-SINR (SC-Max-SINR):} A throughput-centric baseline that greedily assigns each RB to the D2D user with the best channel conditions, subject to MUE interference constraints.
    \item \textbf{Proposed Single-Channel Rate-Fair (SCRF):} A balanced approach designed to bridge the gap between SCRR and SC-Max-SINR, ensuring fair resource distribution while maintaining high spectral efficiency in single-RB scenarios.
    \item \textbf{Multi-Channel Round Robin (MCRR):} Extends SCRR to allow RB aggregation. It employs sequential selection with an Interference Avoidance (IA) algorithm to manage the increased interference resulting from multi-RB usage.
    \item \textbf{Multi-Channel Max-SINR (MC-Max-SINR):} Extends SC-Max-SINR to support multi-RB allocation. It prioritizes users with the highest SINR for each RB, utilizing the IA strategy to filter out users causing severe interference.
    \item \textbf{Proposed Multi-Channel Rate-Fair (MCRF):} Our comprehensive solution that enables shared spectrum reuse. By integrating fairness criteria with the IA strategy, it allows multiple D2D users to multiplex on the same RBs efficiently, maximizing total EE while satisfying fairness requirements.
\end{itemize}

\subsection{Performance Evaluation}
The performance of all compared methods is evaluated using the following metrics: (a) energy efficiency (EE) of D2D users, (b) Jain's fairness index (JFI)~\cite{R19} of D2D users, (c) throughput of D2D users, (d) throughput of MUEs, and (e) system throughput, all measured in the average case. Additionally, the impact of the distance between D2D users and the number of D2D pairs on aforementioned performance metrics is analyzed.

\begin{figure*}[!ht]
	\centering
	\subfigure[EE of D2D users]{
		\label{fig:4-a}
		\includegraphics[width=0.325\textwidth]{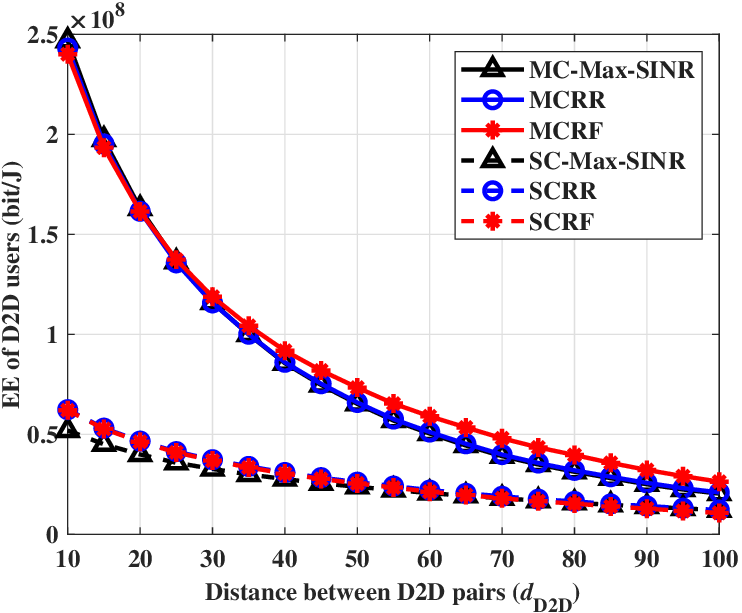}
	}%
	\subfigure[JFI of D2D users]{
		\label{fig:4-b}
		\includegraphics[width=0.325\textwidth]{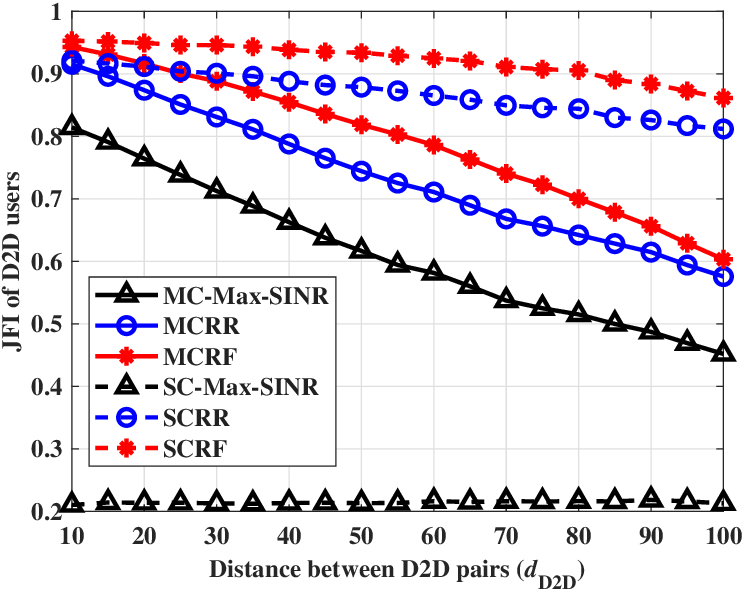}
	}%
	\subfigure[Throughput of D2D users]{
		\label{fig:4-c}
		\includegraphics[width=0.325\textwidth]{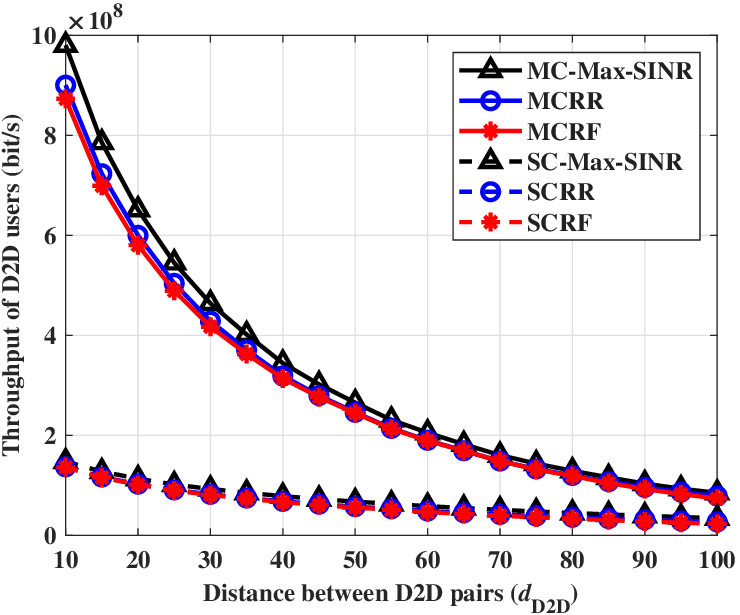}
	}\\
	\subfigure[Throughput of MUEs]{
		\label{fig:4-d}
		\includegraphics[width=0.325\textwidth]{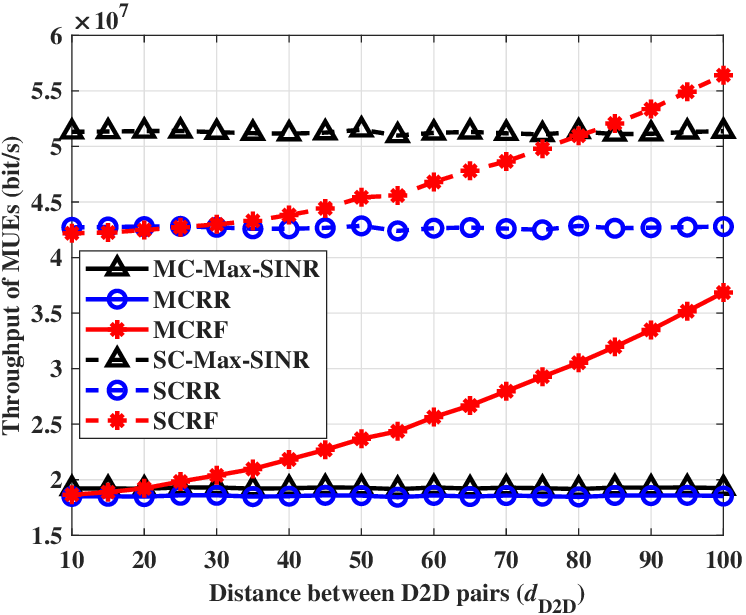}
	}%
	\subfigure[System Throughput]{
		\label{fig:4-e}
		\includegraphics[width=0.325\textwidth]{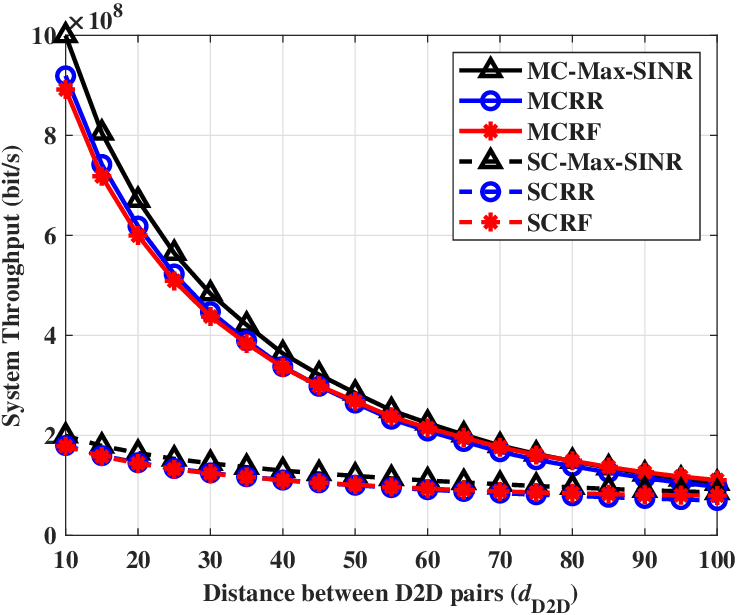}
	}
	\caption{Impact of distance between D2D users on 
		\subref{fig:4-a} EE of D2D users, 
		\subref{fig:4-b} JFI of D2D users, 
		\subref{fig:4-c} Throughput of D2D users, 
		\subref{fig:4-d} Throughput of MUEs, and
		\subref{fig:4-e} System Throughput.}%
	\label{fig:4}
\end{figure*}

\subsubsection{Impact of Distance Between D2D Users}
\label{subsec:impact_d2d_distance}

Fig.~\ref{fig:4} illustrates the system performance as the transmission distance between D2D pairs ($d_{\rm D2D}$) varies from 10 m to 100 m. The analysis and quantitative insights derived from the simulation results are presented as follows:

\begin{itemize}
    \item \textbf{Energy Efficiency and Channel Hardening (Fig.~\ref{fig:4-a}):} 
    As illustrated in Fig.~\ref{fig:4-a}, the Energy Efficiency (EE) decreases monotonically as $d_{\rm D2D}$ increases. This is a direct physical consequence of path loss: larger distances require higher transmit power to support diminishing data rates. Notably, in the single-channel scenario, the proposed SCRF algorithm demonstrates a substantial advantage, outperforming the conventional SCRR scheme by approximately 397\%. In the multi-channel case, MCRF remains competitive with MC-Max-SINR, indicating that it effectively allocates spectral resources only to links that can maintain a viable capacity-to-power ratio.

    \item \textbf{Fairness Resilience (Fig.~\ref{fig:4-b}):} 
    Fig.~\ref{fig:4-b} highlights a stark contrast in fairness performance. As the D2D distance increases, the link quality of edge users degrades, making them vulnerable to starvation. The conventional MC-Max-SINR scheme (black curve) exhibits a steep decline in fairness, dropping drastically to 0.45 at $d_{\rm D2D}=100$ m, as it abandons weak users to maximize sum-rate. In contrast, the proposed MCRF (red curve) maintains a high Fairness Index (JFI $> 0.6$) throughout the entire range. This robustness confirms that the Rate-Fair heuristic successfully prevents the rich-get-richer phenomenon, ensuring equitable access even under poor channel conditions.

    \item \textbf{Throughput Trade-off (Fig.~\ref{fig:4-c}):} 
    The results highlight the benefits of shared reuse. Specifically, at $d_{\rm D2D}=10$ m, the multi-channel MCRF scheme achieves a massive throughput gain of 542\% (reaching 737.1 Mbps) compared to the single-channel SCRF baseline. While MCRF achieves slightly lower throughput than MC-Max-SINR (a marginal gap of $\sim$3.04\% at $d_{\rm D2D}=50$ m), this represents a deliberate design choice. This minor sacrifice yields a significant 7.43\% improvement in fairness, effectively maximizing the aggregate social welfare rather than individual peaks.

    \item \textbf{System Dynamics and MUE Protection Strategy (Figs.~\ref{fig:4-d} \& \ref{fig:4-e}):} 
    The most critical insight appears in the interaction between D2D and MUE throughputs. 
    In the short-range regime ($d_{\rm D2D} < 50$ m), D2D links are strong, and their high capacity dominates the system sum-rate. However, in the long-range regime ($d_{\rm D2D} \ge 80$ m), a regime shift occurs. As D2D link conditions deteriorate, the proposed IA strategy acts as a rigid spectral pruner, automatically revoking resources from failing D2D links. 
    
    \item \textbf{Summary of Insights:} 
    As shown in Fig.~\ref{fig:4-d}, this pruning effect allows the MUE throughput of the MCRF scheme to rise significantly to $\sim$55 Mbps, whereas the Max-SINR scheme keeps MUEs suppressed at a low level of $\sim$18 Mbps by persistently scheduling interfering D2D users.
    This 3-fold improvement in MUE protection allows the total system throughput of MCRF to converge with and eventually surpass that of Max-SINR at $d_{\rm D2D}=100$ m (Fig.~\ref{fig:4-e}). This validates that in scenarios with weak secondary links, prioritizing primary network (MUE) protection yields a higher network-wide spectral efficiency.
\end{itemize}

\begin{figure*}[!ht]
	\centering
	\subfigure[EE of D2D users]{
		\label{fig:5-a}
		\includegraphics[width=0.325\textwidth]{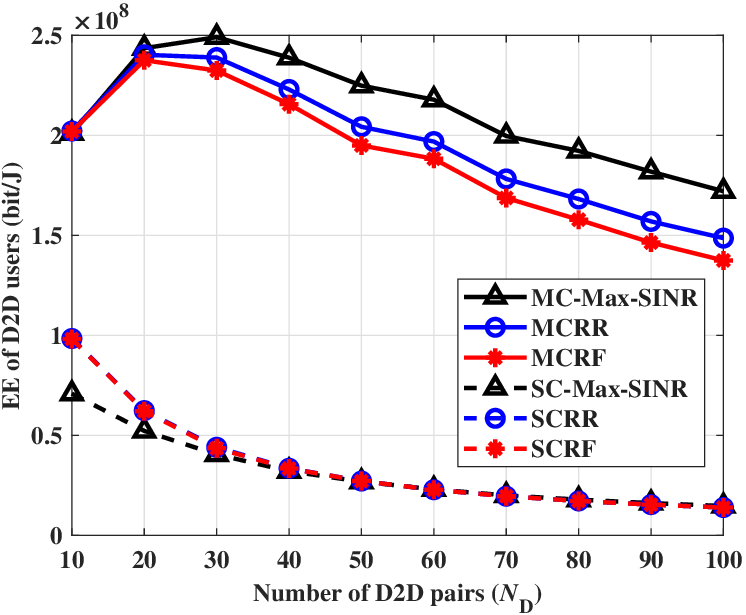}
	}%
	\subfigure[JFI of D2D users]{
		\label{fig:5-b}
		\includegraphics[width=0.325\textwidth]{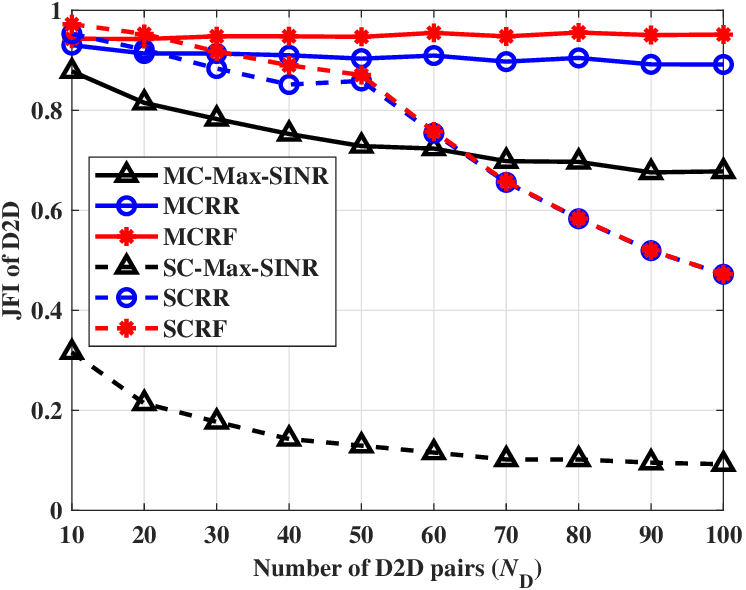}
	}%
	\subfigure[Throughput of D2D users]{
		\label{fig:5-c}
		\includegraphics[width=0.325\textwidth]{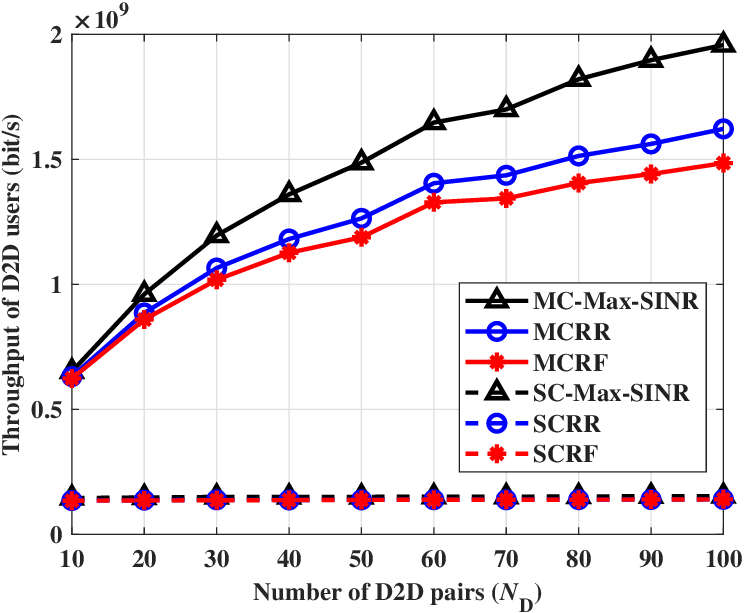}
	}\\
	\subfigure[Throughput of MUEs]{
		\label{fig:5-d}
		\includegraphics[width=0.325\textwidth]{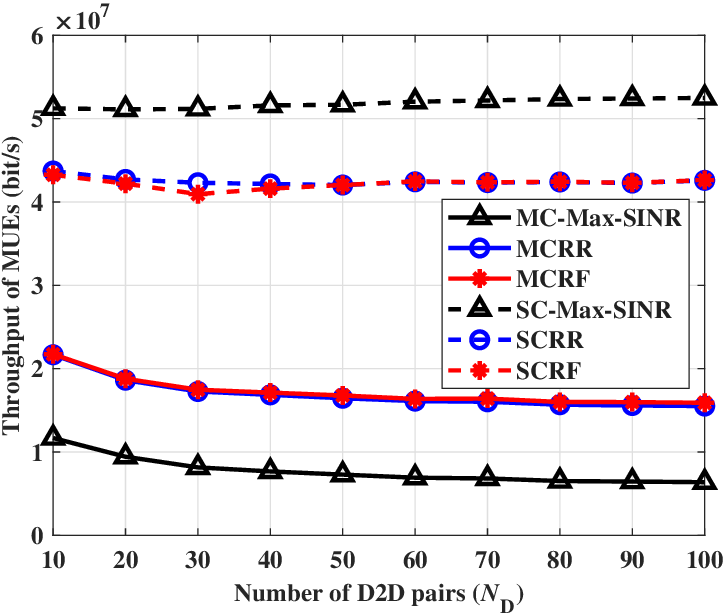}
	}%
	\subfigure[System Throughput]{
		\label{fig:5-e}
		\includegraphics[width=0.33\textwidth]{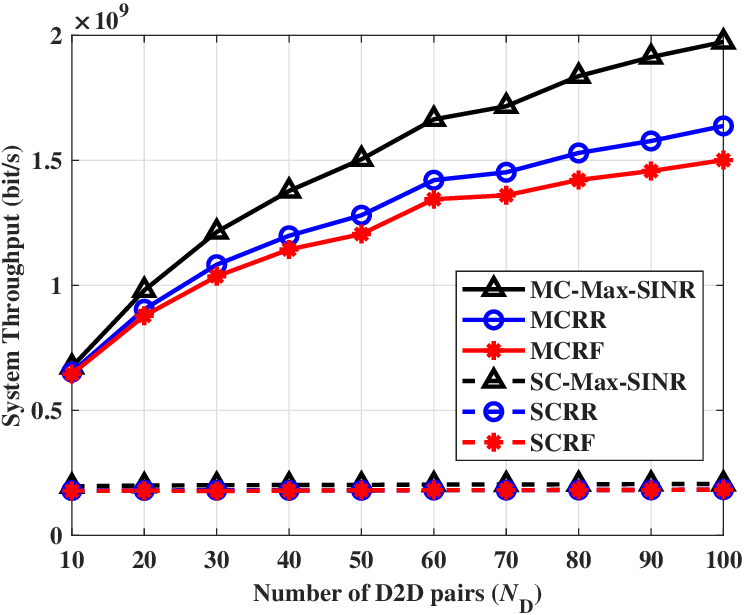}
	}
	\caption{Impact of number of D2D pairs on 
		\subref{fig:5-a} EE of D2D users, 
		\subref{fig:5-b} JFI of D2D users, 
		\subref{fig:5-c} Throughput of D2D users, 
		\subref{fig:5-d} Throughput of MUEs, and
		\subref{fig:5-e} System Throughput.}%
	\label{fig:5}
\end{figure*}

\subsubsection{Impact of Number of D2D Pairs}
\label{subsec:impact_number_d2d}
Fig.~\ref{fig:5} investigates the system performance as the number of D2D pairs ($N_{\rm D}$) varies from 10 to 100, revealing the network's behavior under different congestion levels. The following key observations can be made:

\begin{itemize}
    \item \textbf{Energy Efficiency and Multi-User Diversity (Fig.~\ref{fig:5-a}):} 
    A distinct rise-and-fall trend in Energy Efficiency (EE) is observed. Initially, as $N_{\rm D}$ increases from 10 to 30, the EE improves significantly due to the multi-user diversity gain. However, a saturation point is observed around $N_{\rm D} = 30$. Beyond this point, the system enters an interference-limited regime where aggregate interference grows faster than capacity gains. Additionally, the fixed circuit power consumption of additional users begins to dilute the overall efficiency.
    It is worth noting that MCRF exhibits slightly lower EE compared to MC-Max-SINR. This represents the necessary energy cost to ensure connectivity for cell-edge users, reflecting a trade-off for the superior fairness shown in Fig.~\ref{fig:5-b}.

    \item \textbf{Fairness Robustness against Congestion (Fig.~\ref{fig:5-b}):} 
    As user density increases, resource competition intensifies. While the Fairness Index (JFI) of the Max-SINR scheme collapses rapidly to $<0.1$ at $N_{\rm D}=100$ due to its winner-takes-all nature, the proposed MCRF algorithm demonstrates remarkable resilience, maintaining a JFI of approximately 0.5. This confirms that the Rate-Fair heuristic effectively acts as a load-balancer to prevent resource starvation even when the spectrum is heavily crowded.

    \item \textbf{Scalability of System Capacity (Fig.~\ref{fig:5-c} \& \ref{fig:5-e}):} 
    Fig.~\ref{fig:5-c} shows that D2D throughput scales linearly with $N_{\rm D}$ in multi-channel scenarios, validating the efficacy of the shared reuse policy. By allowing multiple pairs to multiplex on the same RBs, the system capacity is not hard-limited by the number of orthogonal channels ($K$). Although MCRF achieves slightly lower throughput than MCRR, this marginal gap reflects the trade-off made to accommodate weaker users. Consequently, the total system throughput exhibits excellent scalability, supporting massive connectivity demands.

    \item \textbf{MUE Protection Strategy (Fig.~\ref{fig:5-d}):} 
    A critical validation of our Interference Avoidance (IA) strategy is shown in Fig.~\ref{fig:5-d}. Despite the surge in D2D density (from 10 to 100 pairs), the throughput of MUEs remains virtually constant at approximately 42 Mbps for the proposed schemes. This indicates that the IA strategy successfully acts as a spectral gatekeeper by clamping the aggregate interference below the safety threshold $I_{n,k}^{\rm max}$. This ensures that the primary network (MUEs) is effectively insulated from the densification of the secondary D2D network.
\end{itemize}

\vspace{.1em}
\subsection{Discussion and Future Work}
While the proposed MCRF algorithm effectively solves the shared reuse resource allocation problem and ensures fairness, there are certain limitations in our current system model that warrant further investigation. 

Specifically, our simulation framework primarily isolates and evaluates the performance of the resource allocation logic under a simplified channel model. In highly dynamic AGIHNs, specific 3D attributes such as UAV hovering heights, varying UAV densities, and complex LoS and Non-LoS channel variations (e.g., probabilistic air-to-ground pathloss models) can significantly influence the interference patterns. Integrating these sophisticated 3D channel models would make the performance evaluation more comprehensive.

In our future work, we plan to address these open issues by replacing the equal-power heuristic with dynamic power control techniques and integrating deep reinforcement learning agents. This will allow the system to adaptively manage complex 3D LoS/NLoS variations and further advance the practicality of AGIHNs.

\section{Conclusion} 
\label{sec:conclusion}
This study investigates the uplink resource allocation problem in Air-Ground Integrated Heterogeneous Networks (AGIHNs), specifically targeting the challenging shared reuse scenario where multiple D2D pairs simultaneously multiplex on the same resource block. To address the severe intra-tier interference arising from this non-orthogonal sharing, we propose a low-complexity Multi-Channel Rate-Fair (MCRF) algorithm coupled with a heuristic Interference Avoidance (IA) strategy. 

Unlike traditional approaches that rely on exclusive reuse or computationally intensive learning models, the proposed framework achieves a superior balance between Energy Efficiency (EE) and fairness in a training-free manner, making it highly suitable for dynamic UAV environments. Simulation results demonstrate distinct performance advantages depending on the reuse mode. Compared to traditional exclusive reuse schemes, the proposed MCRF algorithm utilizes spectral aggregation to significantly improve D2D Energy Efficiency and Throughput by approximately 397\% and 542\%, respectively. Furthermore, relative to multi-channel benchmarks, it optimizes the efficiency-fairness trade-off, enhancing the Fairness Index by 7.43\%. These findings validate that the proposed strategy effectively unlocks the potential of shared spectrum access in AGIHNs while ensuring robust QoS guarantees for both primary and secondary networks.


\bibliographystyle{IEEEtran}
\bibliography{IEEEabrv,Reference}
\ifCLASSOPTIONcaptionsoff  \newpage \fi

\begin{IEEEbiography}[{\includegraphics[width=1in,height=1.25in,clip,keepaspectratio]{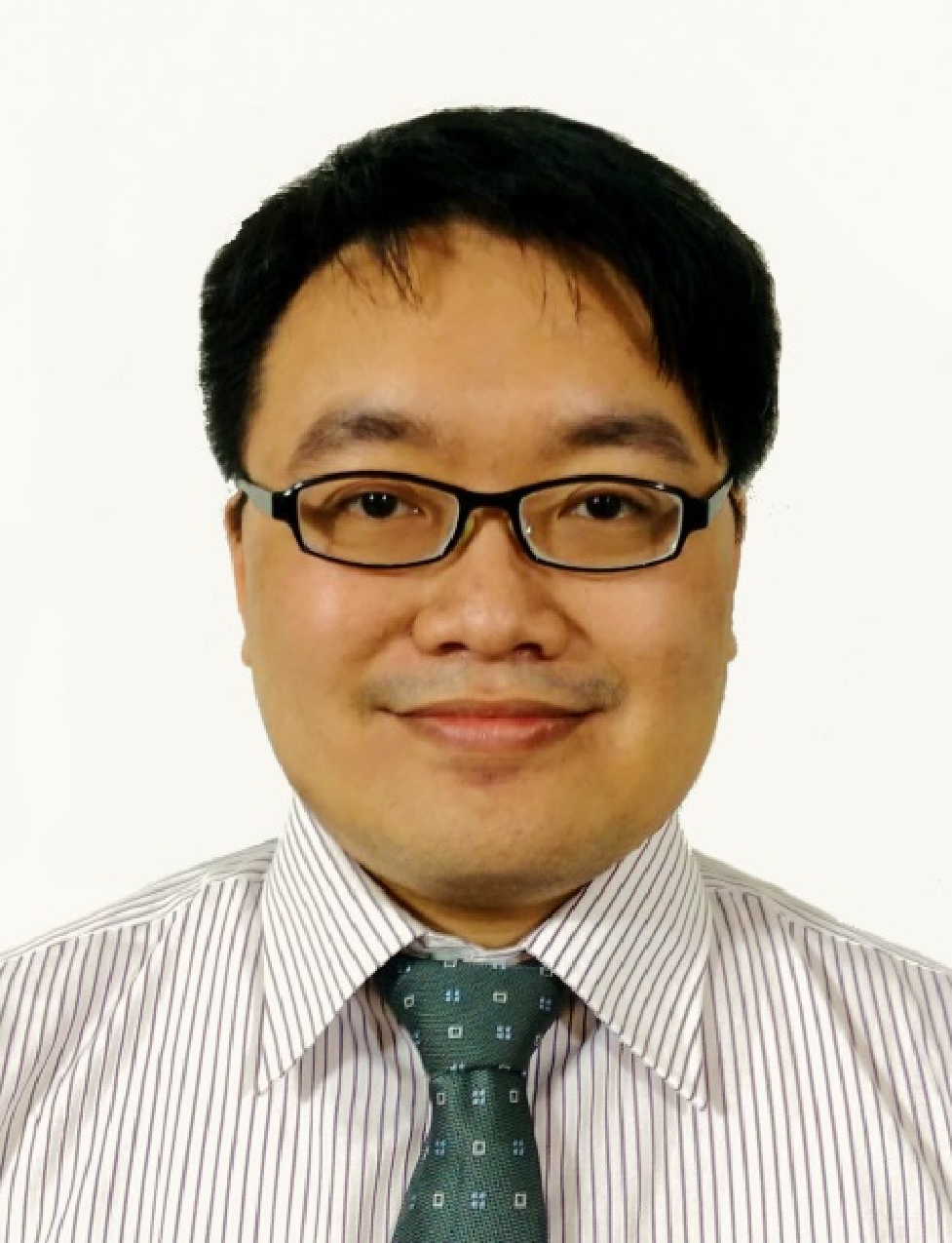}}]{Chuan-Chi Lai}
	(Member, IEEE) received the Ph.D. degree in Computer Science and Information Engineering from the National Taipei University of Technology, Taiwan, in 2017. He held research and faculty positions at National Chiao Tung University and Feng Chia University prior to his current role. Since 2024, he has been an Assistant Professor with the Department of Communications Engineering, National Chung Cheng University, Minxiong Township, Chiayi County, Taiwan. His research interests include mobile edge computing, UAV networks, and AI for wireless communications. Dr. Lai was a recipient of the Postdoctoral Researcher Academic Research Award from the NSTC, Taiwan, in 2019, and Best Paper Awards at WOCC (2018, 2021) and ICUFN (2015).
\end{IEEEbiography}

\begin{IEEEbiography}[{\includegraphics[width=1in,height=1.25in,clip,keepaspectratio]{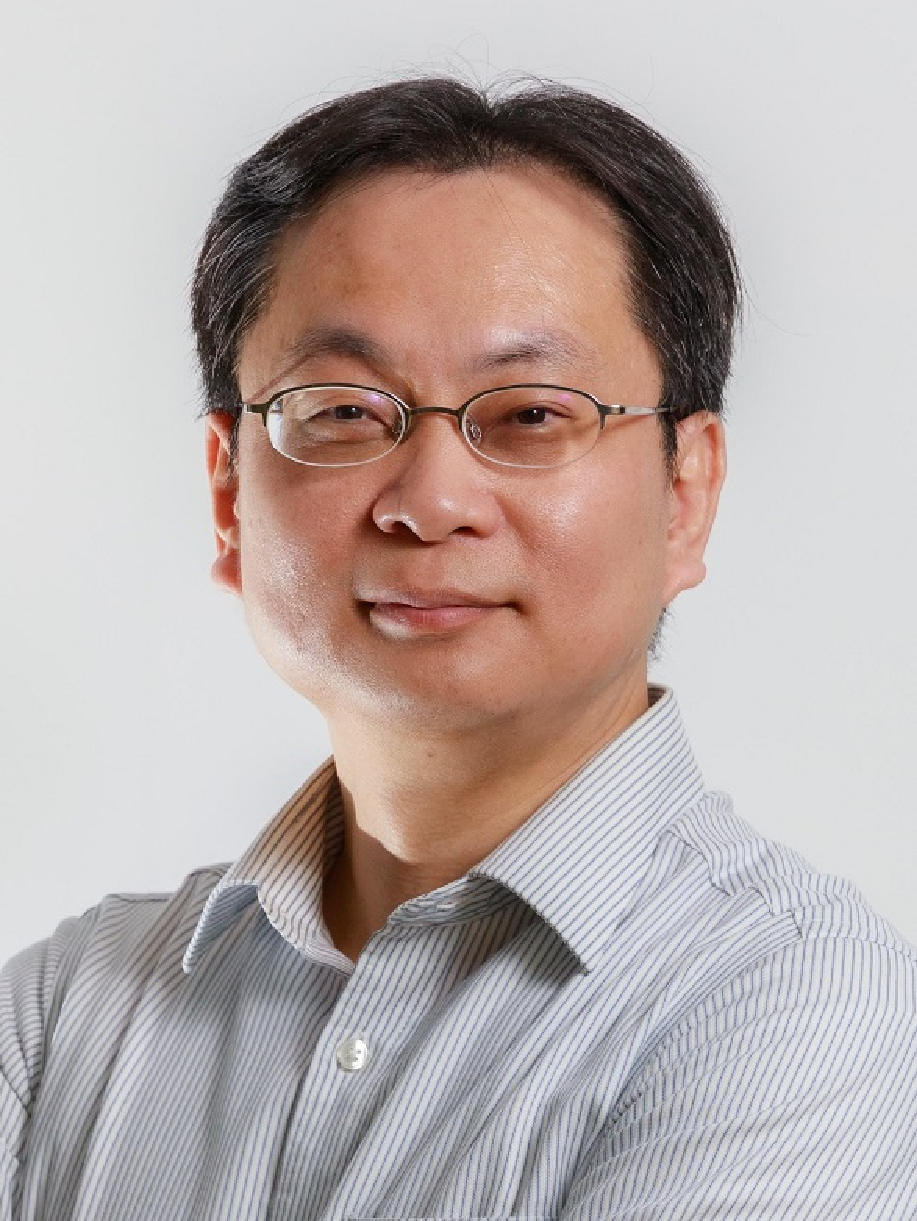}}]{Ang-Hsun Tsai}
	(Member, IEEE) received his Ph.D. Degree in Communication Engineering from National Chiao Tung University, Hsinchu, Taiwan, in 2012. He joined Feng Chia University, Taichung, Taiwan, as an Assistant Professor in 2022. Since 2025, he has been an Associate Professor in the Department of Communications Engineering at Feng Chia University, Taichung, Taiwan. His current research interests include radio resource management in heterogeneous networks, such as 6G mobile networks, non-terrestrial networks, aerial communication networks, and disaster-resilient communication networks. His work also focuses on ubiquitous connectivity, AI-driven communication systems, and integrated sensing and communication technologies.
\end{IEEEbiography}

\begin{IEEEbiography}[{\includegraphics[width=1in,height=1.25in,clip,keepaspectratio]{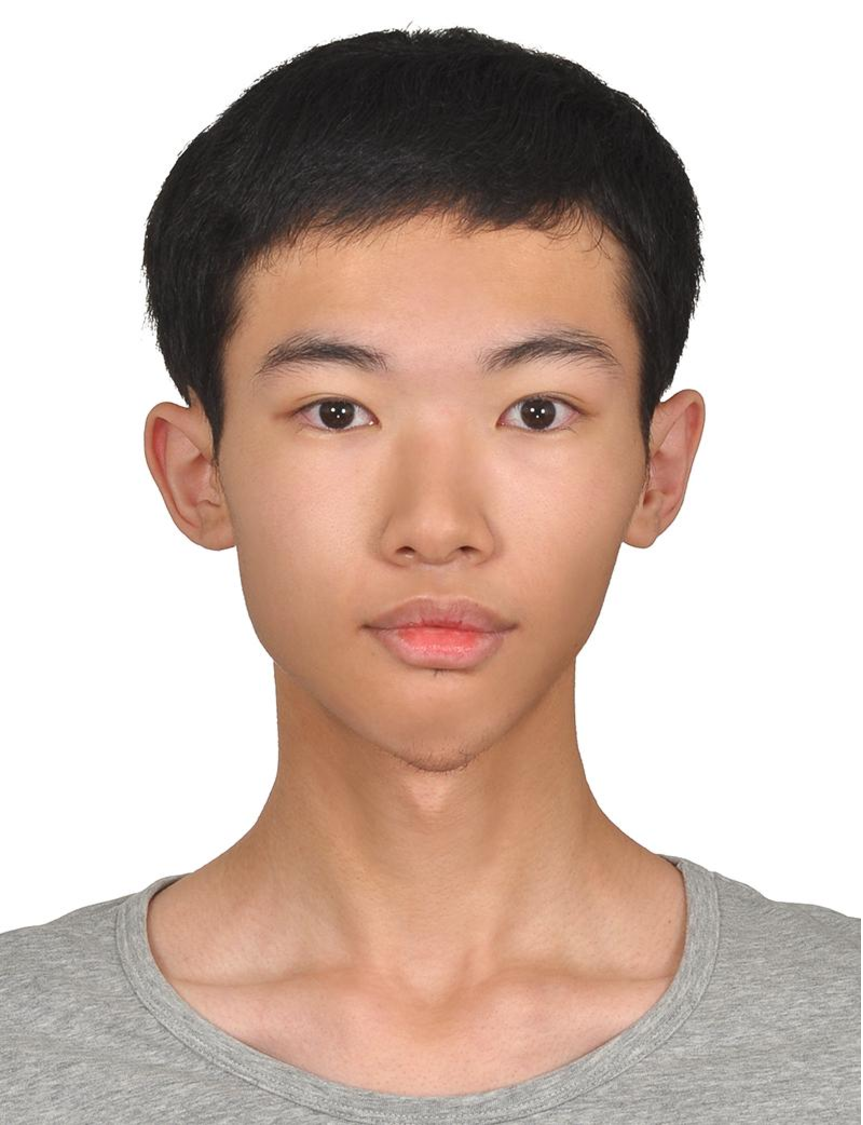}}]{Shang-Long Wu}
	received the B.S. Degree in the Department of Communications Engineering at National Chung Cheng University, Chiayi, Taiwan. He is currently pursuing the M.S. degree with the Institute of Network Engineering, National Yang Ming Chiao Tung University, Hsinchu, Taiwan. His research interests include algorithm design, 6G wireless networks, and AI-driven applications.
\end{IEEEbiography}

\end{document}